%% file: edbt2019.tex
\documentclass[sigconf,edbt]{acmart-edbt2019}

\setcopyright{rightsretained}

\acmDOI{}

\acmISBN{978-3-89318-081-3}

\acmConference[EDBT 2019]{22nd International Conference on Extending Database Technology (EDBT)}{March 26-29, 2019}{Lisbon, Portugal} 
\acmYear{2019}

\usepackage{times,amsmath}
\usepackage{booktabs} 
\usepackage{multirow,tabularx}
\usepackage[ruled,vlined,linesnumbered]{algorithm2e}
\usepackage{paralist}
\usepackage{graphicx}    
\usepackage{amsthm}
\usepackage{xcolor} 
\usepackage[normalem]{ulem} 

\theoremstyle{definition}
\newtheorem{prop}{Proposition}

\begin{document}
\title{MinoanER: Schema-Agnostic, Non-Iterative, Massively
 Parallel Resolution of Web Entities}

\author{Vasilis Efthymiou}
\authornote{Work conducted during the Ph.D research of the author at ICS-FORTH.}
\affiliation{%
  \institution{IBM Research}
}
\email{vasilis.efthymiou@ibm.com}

\author{George Papadakis}
\affiliation{%
  \institution{University of Athens}
}
\email{gpapadis@di.uoa.gr}

\author{Kostas Stefanidis}
\affiliation{%
  \institution{Univeristy of Tampere}
}
\email{konstantinos.stefanidis@tuni.fi}

\author{Vassilis Christophides}
\affiliation{%
  \institution{INRIA-Paris and Univ. of Crete}
}
\email{vassilis.christophides@inria.fr}

\renewcommand{\shortauthors}{}

\begin{abstract}
Entity Resolution (ER) aims to identify different descriptions in various Knowledge Bases (KBs) that refer to the same entity. ER is challenged by the \textit{Variety}, \textit{Volume} and \textit{Veracity} of entity descriptions published in the Web of Data. To address them, we propose the MinoanER framework that simultaneously fulfills \textit{full automation}, support of \textit{highly heterogeneous} entities, and \textit{massive parallelization} of the ER process. MinoanER leverages a token-based similarity of entities to define a new metric that derives the similarity of neighboring entities from the most important relations, as they are indicated only by statistics. A \textit{composite blocking method} is employed to capture different sources of matching evidence from the content, neighbors, or names of entities. The search space of candidate pairs for comparison is compactly abstracted by a novel \textit{disjunctive blocking graph} and processed by a non-iterative, massively parallel matching algorithm that consists of four generic, schema-agnostic matching rules that are quite robust with respect to their internal configuration. We demonstrate that the effectiveness of MinoanER is comparable to existing ER tools over real KBs exhibiting \textit{low Variety}, but it outperforms them significantly when matching KBs with \textit{high Variety}.
\end{abstract}

%
%


\maketitle

\input{1_intro}

\input{2_preliminaries}
\input{3_blockingGraph}
\input{4_matching}
\input{5_relatedWork}
\input{6_experiments}

\input{7_conclusion}

\bibliographystyle{ACM-Reference-Format}
\bibliography{bibliography}

\end{document}

%% file: 1_intro.tex
\section{Introduction}\label{sec:matching_intro}
Even when data integrated from multiple sources refer to the same real-world entities (e.g., persons, places), they usually exhibit several quality issues such as \emph{incompleteness} (i.e., partial data), \emph{redundancy} (i.e., overlapping data), \emph{inconsistency} (i.e., conflicting data) or simply \emph{incorrectness} (i.e., data errors). A typical task for improving various data quality aspects is \emph{Entity Resolution} (ER). In the Web of Data, ER aims to facilitate interlinking of data that describe the same real-world entity, when unique entity identifiers are not shared across different Knowledge~Bases~(KBs) describing them \cite{DBLP:series/synthesis/2015Christophides}. To resolve entity descriptions we need (a) \emph{to compute effectively the similarity of entities}, and (b) \emph{to pair-wise compare entity descriptions}. 
Both problems are challenged by the three Vs of the Web of Data, namely Variety, Volume and Veracity~\cite{DBLP:series/synthesis/2015Dong}. Not only does the \emph{number of entity descriptions} published by each KB never cease to increase, but also the \emph{number of KBs} even for a single domain, has grown to thousands (e.g., there is a x100 growth of the LOD cloud size since its first edition\footnote{\label{fn:lod}\url{https://lod-cloud.net}}). Even in the same domain, KBs are \emph{extremely heterogeneous} both regarding how they semantically structure their data, as well as how diverse properties are used to describe even substantially similar entities (e.g., only 109 out of $\sim$2,600 LOD vocabularies are shared by more than one KB). Finally, KBs are of \emph{widely differing quality}, with significant differences in the coverage, accuracy and timeliness of data provided \cite{DBLP:journals/semweb/Debattista0AC18}. Even in the same domain, various inconsistencies and errors in entity descriptions may arise, due to the limitations of the automatic extraction tools \cite{Wang:2015:DXD:2723372.2750549}, or of the crowd-sourced contributions.

\begin{figure}[t]
	\center \includegraphics[width=0.95\columnwidth]{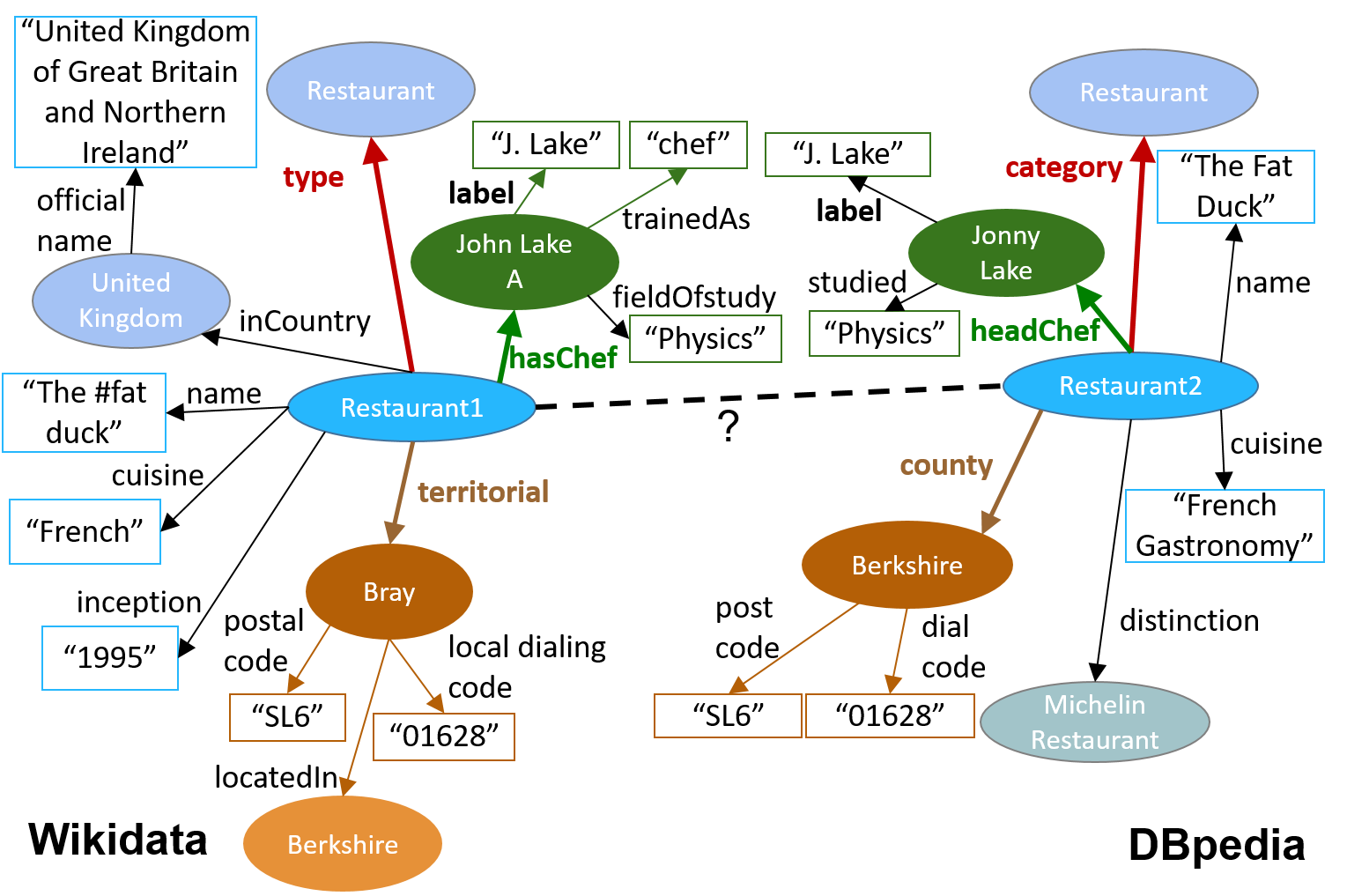}
	\caption{Parts of entity graphs, representing the Wikidata (left) and DBpedia (right) KBs.}
	\label{fig:entity_graph}
	\vspace{-10pt}
\end{figure}

The Web of Data essentially calls for novel ER solutions that relax a number of assumptions underlying state-of-the-art methods. The most important one is related to the notion of similarity that better characterizes \textit{entity descriptions} in the Web of Data - we define an entity description to be a URI-identifiable set of attribute-value pairs, where values can be literals, or the URIs of other descriptions, this way forming an \emph{entity graph}. Clearly, Variety results in extreme \textit{schema heterogeneity}, with an unprecedented number of attribute names that cannot be unified under a global schema \cite{DBLP:conf/pods/GolshanHMT17}. This situation renders all schema-based similarity measures that compare specific attribute values inapplicable~\cite{DBLP:conf/pods/GolshanHMT17}. We thus argue that similarity evidence of entities within and across KBs can be obtained by looking at the bag of strings contained in descriptions, regardless of the corresponding attributes. As this \textbf{value-based similarity} of entity pairs may still be weak, due to a highly heterogeneous description content, we need to consider additional sources of matching evidence; for instance, the \textbf{similarity of neighboring entities}, which are interlinked via various semantic relations. 

Figure~\ref{fig:entity_graph} presents parts of the Wikidata and DBpedia KBs, showing the entity graph that captures connections inside them. For example, Restaurant2 and Jonny Lake are neighbor entities in this graph, connected via a ``headChef'' relation. 
If we compare John Lake A to Jonny Lake based on their values only, it is easy to infer that those descriptions are matching; they are \emph{strongly similar}. However, we cannot be that sure about Restaurant1 and Restaurant2, if we only look at their values. Those descriptions are \textit{nearly similar} and we have to look further at the similarity of their neighbors (e.g, John Lake A and Jonny Lake) to verify that they match. 

\begin{figure}[t]
	\center \includegraphics[width=0.9\columnwidth]{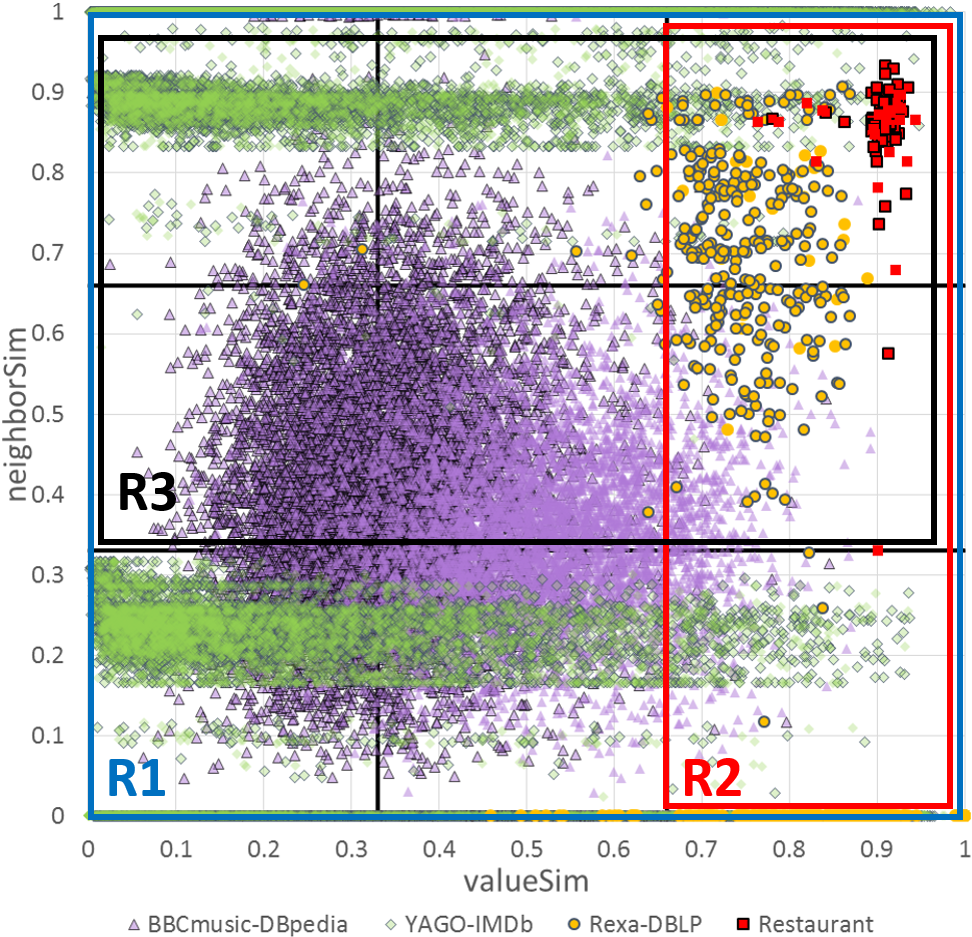}
	\caption{Value and neighbor similarity distribution of matches in 4	datasets (see Table \ref{tab:matching_datasets} for more details).}
    \vspace{-15pt}
	\label{fig:similarityDistributions}
\end{figure}

Figure~\ref{fig:similarityDistributions} depicts both sources of similarity evidence ({\sl valueSim}, {\sl neighborSim}) for entities known to match (i.e., ground truth) in four benchmark datasets that are frequently used in the literature (details in Table~\ref{tab:matching_datasets}). Every dot corresponds to a matching pair of entities, and its shape denotes its origin KBs. The horizontal axis reports the normalized value similarity (weighted Jaccard coefficient~\cite{DBLP:conf/kdd/Lacoste-JulienPDKGG13}) based on the tokens (i.e., single words in attribute values) shared by a pair of descriptions, while the vertical one reports the maximum value similarity of their neighbors. The value similarity of matching entities significantly varies across different KBs. For \textbf{strongly similar entities}, e.g., with a value similarity $>$ 0.5, existing duplicate detection techniques work well. However, a large part of the matching pairs of entities is covered by \textbf{nearly
similar entities}, e.g., with a value similarity $<$ 0.5. To resolve them, we need to additionally exploit evidence regarding the similarity of neighboring entities.

This also requires revisiting the blocking (aka indexing) techniques used to reduce the number of candidate pairs~\cite{Christen11}. To avoid restricting candidate matches (i.e., descriptions placed on the same block) to strongly similar entities, we need to assess both \textit{value} and \textit{neighbor} similarity of candidate matches. In essence, rather than a unique indexing function, we need to consider a \emph{composite blocking} that provides matching evidence from different sources, such as the content, the neighbors or even the names (e.g., {\small \texttt{rdfs:label}}) of entities. Creating massively parallelizable techniques for 
processing the search space of candidate pairs formed by such composite blocking is an open research challenge.

Overall, the main requirements for Web-scale ER are: \emph{(i)}
identify both strongly and nearly similar matches, \emph{(ii)} do not rely on a given schema, \emph{(iii)} do not rely on domain experts for aligning relations and matching rules, \emph{(iv)}
develop non-iterative solutions to avoid late convergence, and \emph{(v)} scale to massive volumes~of~data. None of the existing ER frameworks proposed for the Web of Data (e.g., 
LINDA~\cite{DBLP:conf/cikm/BohmMNW12}, SiGMa~\cite{DBLP:conf/kdd/Lacoste-JulienPDKGG13} and
RiMOM~\cite{DBLP:journals/jcst/ShaoHLWCX16}) simultaneously fulfills all these requirements. 
In this work, we present the MinoanER framework for a Web-scale ER\footnote{A preliminary, abridged version of this paper appeared in \cite{icdeEfthymiou2018}.}. 
More precisely, we make the following contributions:

$\bullet$ 
We leverage a token-based similarity of entity descriptions, introduced in \cite{DBLP:journals/tkde/PapadakisKPN14}, to define a new metric for the similarity of a set of neighboring entity pairs that are linked via important relations to the entities of a candidate pair. Rather than requiring an {\it a priori} knowledge of the entity types or of their correspondences, we rely on \emph{simple statistics over two KBs to recognize the most important entity relations} involved in their neighborhood, as well as, \emph{the most distinctive attributes that could serve as names of entities} beyond the {\small \texttt{rdfs:label}}, which is not always available in descriptions. 

$\bullet$ 
We exploit several indexing functions to place entity descriptions in the same block either because they share a common token in their values, or they share a common name. Then, we introduce a novel abstraction of multiple sources of matching evidence regarding
a pair of entities (from the content, neighbors, or the names of their descriptions) under the form of a \emph{disjunctive blocking graph}. We present an efficient algorithm for weighting and then pruning the edges with low weights, which are unlikely to correspond to matches. As opposed to existing disjunctive blocking schemes~\cite{DBLP:conf/icdm/BilenkoKM06,DBLP:journals/corr/Kejriwal16}, \emph{our disjunctive blocking is schema-agnostic and requires no (semi-)supervised learning}.

$\bullet$ 
We propose a \textit{non-iterative matching process} that is implemented in Spark~\cite{DBLP:conf/hotcloud/ZahariaCFSS10}. Unlike the data-driven convergence of existing systems (e.g., LINDA~\cite{DBLP:conf/cikm/BohmMNW12}, SiGMa~\cite{DBLP:conf/kdd/Lacoste-JulienPDKGG13}, RiMOM~\cite{DBLP:journals/jcst/ShaoHLWCX16}), the matching process of MinoanER involves a specific number of predefined generic, schema-agnostic matching rules (\textsf{R1}-\textsf{R4}) that traverse the blocking graph. First, we identify matches based on their name (\textsf{R1}). This is a very effective method that can be applied to all descriptions, regardless of their values or neighbor similarity. Unlike the schema-based blocking keys of relational descriptions usually provided by domain experts, \emph{MinoanER automatically specifies distinctive names of entities from data statistics}. Then, the value similarity is exploited to find matches with many common and infrequent tokens, i.e., strongly similar matches (\textsf{R2}). When value similarity is not high, nearly similar matches are identified based on both value and neighbors' similarity using a \emph{threshold-free rank aggregation function} (\textsf{R3}), as opposed to existing works that combine different matching evidence into an aggregated score. Finally, \emph{reciprocal evidence} of matching is exploited as a verification of the returned results: only entities that are mutually ranked in the top positions of their unified ranking lists are considered matches (\textsf{R4}).  Figure \ref{fig:similarityDistributions} abstractly illustrates the type of matching pairs that are covered~by~each~matching rule.

$\bullet$ 
We experimentally compare the effectiveness of MinoanER against state-of-the-art methods using established benchmark data that involve real KBs. The main conclusion drawn from our experiments is that MinoanER achieves at least equivalent performance over KBs exhibiting a \textit{low Variety} (e.g., those originating from a common data source like Wikipedia) even though the latter make more assumptions about the input KBs (e.g., alignment of relations); yet, MinoanER significantly outperforms state-of-the-art ER tools when matching KBs with high Variety. The source code and datasets used in our experimental study are publicly available\footnote{\url{http://csd.uoc.gr/~vefthym/minoanER}}.

The rest of the paper is structured as follows: we introduce our value and neighbor similarities in Section~\ref{sec:matching_preliminaries}, and we delve into the blocking schemes and the blocking graph that lie at the core of our approach in Section \ref{sec:matching_blocking}. Section~\ref{sec:matching_matching} describes the matching rules of our approach along with their implementation in Spark, while
Section~\ref{sec:matching_relatedWork} overviews the main differences with the state-of-the-art ER methods. We present our thorough experimental analysis in Section~\ref{sec:matching_experiments} and 
we conclude the paper in Section~\ref{sec:matching_conclusions}.

%% file: 2_preliminaries.tex
\section{Basic Definitions}\label{sec:matching_preliminaries}

Given a KB $\mathcal{E}$, an entity description with a URI identifier $i$, denoted by $e_i \in \mathcal{E}$, is a set of attribute-value pairs about a real-world entity.
When the identifier of an entity description $e_j$ appears in the values of another entity description $e_i$, the corresponding attribute is called a \textit{relation} and the corresponding value ($e_j$) a \textit{neighbor} of $e_i$. More formally, the relations of $e_i$ 
are defined as $relations(e_i) = \{p | (p, j) \in e_i \wedge e_j \in \mathcal{E}\}$, while its neighbors as $neighbors(e_i) = \{e_j | (p, j) \in e_i \wedge e_j \in \mathcal{E}\}$. 
For example, for the Wikidata KB in the left side of Figure~\ref{fig:entity_graph} we have: $relations(Restaurant1)$ = \{hasChef, territorial, inCountry\}, and $neighbors(Restaurant1)$ = \{John Lake A, Bray, United Kingdom\}.

In the following, we exclusively consider clean-clean ER, i.e., the
sub-problem of ER that seeks matches among two 
duplicate-free (clean) KBs.
Thus, we simplify the presentation of our approach, but 
the proposed techniques can be easily generalized to more than two clean KBs or a single dirty KB, i.e., a KB that contains duplicates.



\subsection{Entity similarity based on values}
Traditionally, similarity between entities is computed based on their values.
In our work, we apply a similarity measure based on the number and the frequency of common words between two values\footnote{We handle numbers and dates in the same way as strings, assuming string-dominated entities.}.



\begin{definition}
\label{def:valuesim}
Given two KBs,  $\mathcal{E}_1$ and $\mathcal{E}_2$,
the \textbf{value similarity} of two entity descriptions $e_i \in \mathcal{E}_1,
e_j \in\mathcal{E}_2$ is defined as: 
$valueSim(e_i,e_j){=}\sum_{t \in tokens(e_i) \cap tokens(e_j)} \frac{1}
{log_2(EF_{\mathcal{E}_1}(t) \cdot EF_{\mathcal{E}_2}(t) + 1)}$,
where $EF_{\mathcal{E}}(t) = |\{e_l | e_l \in \mathcal{E} \wedge t \in
tokens(e_l)\}|$ stands for ``Entity Frequency'', which is the number of entity
descriptions in $\mathcal{E}$ having token $t$ in their values.
\end{definition}

This value similarity shares the same intuition as TF-IDF in information retrieval. 
If two entities share many, infrequent tokens, then they have high value similarity. 
On the contrary, very frequent words (resembling stopwords in information retrieval) are not considered an important matching evidence, when they are shared by two descriptions, and therefore, they only contribute insignificantly to the $valueSim$ score. 
The number of common words is accounted by the number of terms that are considered in the sum and the frequency of those words is given by the inverse Entity Frequency (EF), similar to the inverse Document Frequency (DF) in information retrieval. 

\begin{prop}\label{prop:metric}
$valueSim$ is a \emph{similarity metric}, since it satisfies the following properties~\cite{DBLP:journals/tcs/ChenMZ09}: \\
$\bullet ~ valueSim(e_i,e_i) \geq 0$, \\
$\bullet ~ valueSim(e_i,e_j) = valueSim(e_j,e_i)$, \\
$\bullet ~ valueSim(e_i,e_i) \geq valueSim(e_i,e_j)$, \\
$\bullet ~ valueSim(e_i,e_i)= valueSim(e_j,e_j)= valueSim(e_i,e_j)$ ${\Leftrightarrow} e_i{=}e_j$, \\
$\bullet ~ valueSim(e_i,e_j) + valueSim(e_j,e_z) \leq valueSim(e_i,e_z)$ $+$\\ $valueSim(e_j,e_j)$.
\end{prop}

\begin{proof}
Please refer to the extended version of this paper\footnote{\url{http://csd.uoc.gr/~vefthym/DissertationEfthymiou.pdf}}. 
\end{proof}


Note that $valueSim$ has the following properties: \emph{(i)} it is not a
normalized metric, since it can take any value in $[0,+\infty)$, with 0
denoting the existence of no common tokens in the values of the compared descriptions.
\emph{(ii)} The maximum contribution of a single common token between two
descriptions is 1, in 
case this common token does not appear in the values
of any other entity description, i.e., when $EF_{\mathcal{E}_1}(t) \cdot EF_{\mathcal{E}_2}(t) = 1$.
\emph{(iii)} It is a \textit{schema-agnostic similarity metric}, as it 
disregards any schematic information\footnote{Note that our value similarity
metric is crafted for the token-level noise in literal values, rather than
the character-level one. Yet, our overall approach is tolerant to
character-level noise, as verified by our extensive experimental analysis with
real datasets that include it. The reason is that it is highly unlikely for
matching entities to have character-level noise in all their common tokens.}.

\subsection{Entity similarity based on neighbors}\label{ssec:neighborsSim}
In addition to value similarity, we exploit the relations between descriptions to find the matching entities of the compared 
KBs.
This can be done by aggregating the value similarity of all pairs of descriptions that are neighbors of the target descriptions.  

Given the potentially high number of neighbors that a description might have, we propose considering only the most valuable neighbors for computing the neighbor similarity between two target descriptions.
These are neighbors that are connected with the target descriptions via important relations, i.e., relations that exhibit high \emph{support} and \emph{discriminability}. Intuitively, high support for a particular relation $p$ indicates that 
$p$ appears in many entity descriptions, 
while high discriminability for $p$ indicates that it has many distinct values. More formally: 

\begin{definition}
The \textbf{support of a relation} $p \in \mathcal{P}$ in a KB 
$\mathcal{E}$ is defined as: 
	$support(p) = \frac{|instances(p)|}{|\mathcal{E}|^2}$, 
	where $instances(p) = \{(i, j) | e_i, e_j \in \mathcal{E} \wedge (p,j) \in
	e_i\}$.
\end{definition}

\begin{definition}
The \textbf{discriminability of a relation} $p \in \mathcal{P}$ in a KB 
$\mathcal{E}$ is defined as:
	$discriminability(p) = \frac{|objects(p)|}{|instances(p)|}$, 
	where $objects(p) = \{j | (i,j) \in instances(p)\}$.
\end{definition} 

Overall, we combine support and discriminability via their harmonic mean 
in order to locate the most important relations. 
\begin{definition}
The \textbf{importance of a relation} $p \in \mathcal{P}$ in a KB 
$\mathcal{E}$ is defined as:	
	$importance(p) = 2\cdot\frac{support(p) \cdot discriminability(p)}{support(p) + discriminability(p)}.$
\end{definition}

On this basis, 
we identify the most valuable relations and neighbors for every single entity description (i.e., \textit{locally}). 
We use $topNrelations(e_i)$ to denote the $N$ relations in $relations(e_i)$ with
the maximum importance scores. Then, the \textbf{best neighbors} for $e_i$ are defined as:
$topNneighbors(e_i) = \{ne_i | (p,ne_i) \in e_i \wedge
p \in topNrelations(e_i)\}$. 

Intuitively, strong matching evidence (high value similarity) for the important neighbors leads to strong matching evidence for the target pair of descriptions. Hence, we formally define neighbor similarity as follows: 

\begin{definition}
\label{def:neighborsim}
Given two KBs, $\mathcal{E}_1$ and $\mathcal{E}_2$, the \textbf{neighbor similarity} of two entity descriptions $e_i \in
\mathcal{E}_1, e_j \in\mathcal{E}_2$ is: 
$$neighborNSim(e_i,e_j){=}\sum\limits_{\substack{ne_i \in topNneighbors(e_i)\\ ne_j \in topNneighbors(e_j)}} valueSim(ne_i, ne_j).$$ 
\end{definition}

\begin{prop}\label{prop:neighbor_metric}
$neighborNSim$ is a similarity metric. 
\end{prop}

\begin{proof}
Given that $neighborNSim$ is the sum of similarity metrics ($valueSim$), it is a similarity metric, too~\cite{DBLP:journals/tcs/ChenMZ09}.
\end{proof}

Neither $valueSim$ nor $neighborNSim$ are normalized, since the number of terms that contribute in the sums is an important matching evidence that can be mitigated if the values were normalized.


\begin{example}
Continuing our example in Figure~\ref{fig:entity_graph}, assume that the best two relations for $Restaurant1$ and $Restaurant2$ are: $top2relations(Restaurant1)$ = \{hasChef, territorial\} and \\
$top2relations(Restaurant2)$ = \{headChef, county\}. 
Then, \\$top2neighbors(Restaurant1)$ = \{John Lake A, Bray\} and \\
$top2neighbors(Restaurant2)$ = \{Jonny Lake, Berkshire\}, and \\
$neighbor2Sim(Restaurant1, Restaurant2)$ = \\
$valueSim(Bray,$ $Jonny Lake)$+$valueSim(John~Lake~A,$ $Berkshire)$ +$valueSim(Bray,$ $Berkshire)$+$valueSim(John~Lake~A,$ $Jonny~Lake)$. 
Note that since we don't have a relation mapping, we also consider the comparisons (Bray, JonnyLake) and (John Lake A, Berkshire).
\end{example}

\textbf{Entity Names.}
From every KB, we also derive the \textit{global} top-$k$ attributes of highest
importance, whose \emph{literal} values act as \textbf{names} for any description $e_i$ that contains them. Their support is simply defined as: $support(p) = |subjects(p)|/|\mathcal{E}|$, where $subjects(p) = \{i | (i,j) \in instances(p)\}$ \cite{DBLP:conf/semweb/SongH11}.
Based on these statistics, function $name(e_i)$ returns the names of $e_i$, and $\mathcal{N}_x$ denotes all names in a KB $\mathcal{E}_x$. In combination with $topNneighbors(e_i)$, this function covers both local and global property importance, exploiting both the rare and frequent attributes that are distinctive enough to designate matching entities.


%% file: 3_blockingGraph.tex
\section{Blocking}\label{sec:matching_blocking}
To enhance performance, \emph{blocking} is typically used as a pre-processing step for ER in order to reduce the number of unnecessary comparisons, i.e., comparisons between descriptions that do not match. After blocking, each description can be compared only to others placed within the same block. The desiderata of blocking are \cite{DBLP:books/daglib/0030287}: $(i)$ to place matching descriptions in common blocks (\emph{effectiveness}), and $(ii)$ to minimize the number of suggested comparisons (\emph{time efficiency}). However, efficiency dictates skipping many comparisons, possibly yielding many missed matches, which in turn implies low effectiveness. Thus, the main objective of blocking is to achieve a good trade-off between minimizing the number of suggested comparisons and minimizing the number of missed matches \cite{Christen11}.

In general, blocking methods are defined over key values that can be used to decide whether or not an entity description could be placed in a block using an \emph{indexing function} \cite{Christen11}. The `uniqueness' of key values determines the number of entity descriptions placed in the same block, i.e., which are  considered as \emph{candidate matches}. 
More formally, the building blocks of a blocking method can be defined as~\cite{DBLP:conf/icdm/BilenkoKM06}:


$\bullet$ An \textit{indexing function} $h_{key} : \mathcal{E} \rightarrow 2^B$ is a unary function that, when applied to an entity description using a specific blocking key, it returns as a value the subset of the set of all blocks $B$, under which the description will be indexed. 

$\bullet$ A \textit{co-occurrence function} $o_{key} : \mathcal{E} \times
\mathcal{E} \rightarrow \{true, false\}$ is a binary function that, when applied to a pair of entity descriptions, it returns `true' if the intersection of the sets of blocks produced by the indexing function on its arguments, is non-empty, and `false' otherwise; $o_{key}(e_k, e_l) = true$ iff $h_{key}(e_k) \cap h_{key}(e_l) \neq \emptyset$.

In this context, each pair of descriptions whose co-occurrence function returns `true' shares at least one common block, and the distinct union of the block elements is the input entity collection (i.e., all the descriptions from a set of input KBs).  Formally: 

\begin{definition}
\label{def:blocking}
Given an entity collection $\mathcal{E}$, \textbf{atomic blocking} is defined by an indexing function $h_{key}$ for which the generated blocks, $B^{key}$=$\{b_1^{key}, \ldots,$ $b_m^{key}\}$, 
satisfy the following conditions:
\begin{enumerate}[(i)]
	\item $\forall e_k, e_l \in b_i^{key} : b_i^{key} \in B^{key}, o_{key}(e_k,e_l) = true$, 
	\item $\forall (e_k,e_l) : o_{key}(e_k,e_l){=}true, \exists b_i^{key} \in B^{key}, e_k, e_l \in b_i^{key}$,
	\item $\bigcup\limits_{b_i^{key} \in B^{key}} b_i^{key} = \mathcal{E}$.
\end{enumerate}
\end{definition}

Given that a single key is not enough for indexing loosely structured and highly heterogeneous entity descriptions, we need to consider several keys that the indexing function will exploit to build different sets of blocks. Such a composite blocking method is characterized by a disjunctive co-occurrence function over the atomic blocks, and it is formally defined as: 

\begin{definition}
\label{def:disjunctiveBlocking}
Given an entity collection $\mathcal{E}$, 
\textbf{disjunctive blocking} is defined by a set of indexing functions $H$, for which the generated blocks $B = \bigcup\limits_{h_{key} \in H}B^{key}$ 
satisfy the following conditions:
\begin{enumerate}[(i)]
	\item $\forall e_k, e_l \in b : b \in B, o_H(e_k,e_l) = true$, 
	\item $\forall (e_k,e_l) : o_H(e_k,e_l) = true, \exists b \in B, e_k, e_l \in b$, 
\end{enumerate}
where $o_H(e_k,e_l) = \bigvee_{h_{key} \in H} o_{key}(e_k,e_l)$.
\end{definition}

Atomic blocking can be seen as a special case of composite blocking, consisting of a singleton set, i.e., $H = \{h_{key}\}$. 

\subsection{Composite Blocking Scheme} 
To achieve a good trade-off between effectiveness and efficiency, our
\textit{composite blocking scheme} assesses the name 
and value similarity of the candidate matches in combination with similarity evidence provided by their neighbors on important relations. We consider the blocks constructed for all entities $e_i \in \mathcal{E}$ using the indexing function $h_i(\cdot)$ both over entity names ($\forall n_j \in names(e_i): h_N(n_j)$) and tokens ($\forall t_j \in tokens(e_i): h_T(t_j)$). The composite blocking scheme $\mathbf{\mathcal{O}}$ of MinoanER is defined by the following disjunctive co-occurrence condition of any two entities $e_i, e_j \in \mathcal{E}$: 
$\mathcal{O}(e_i, e_j) = o_N(e_i, e_j) \vee o_T(e_i, e_j) \vee $ \\ $(\bigvee_{(e'_{i},e'_{j}) \in topNneighbors(e_i) \times topNneighbors(e_j)} o_T(e'_{i},e'_{j})),$ 
where $o_N$, $o_T$ is the co-occurrence function applied on names and tokens, respectively. Intuitively, two entities are placed in a common block, and are then considered candidate matches, if at least one of the following three cases holds: $(i)$ they have the same name, which is not used by any other entity, in which case the common block contains only those two entities, or $(ii)$ they have at least one common word in any of their values, in which case the size of the common block is given by the product of the Entity Frequency ($EF$) of the common term in the two input collections, or
$(iii)$ their top neighbors share a common word in any of their values. Note that token blocking (i.e., $h_T$) allows for deriving $valueSim$ from the size of blocks shared by two descriptions. As a result, no additional blocks are needed to assess neighbor similarity of candidate entities: token blocking is sufficient also for estimating $neighborNsim$ according to Definition~\ref{def:neighborsim}.

\begin{figure}[t]
	\center \includegraphics[width=0.6\columnwidth]{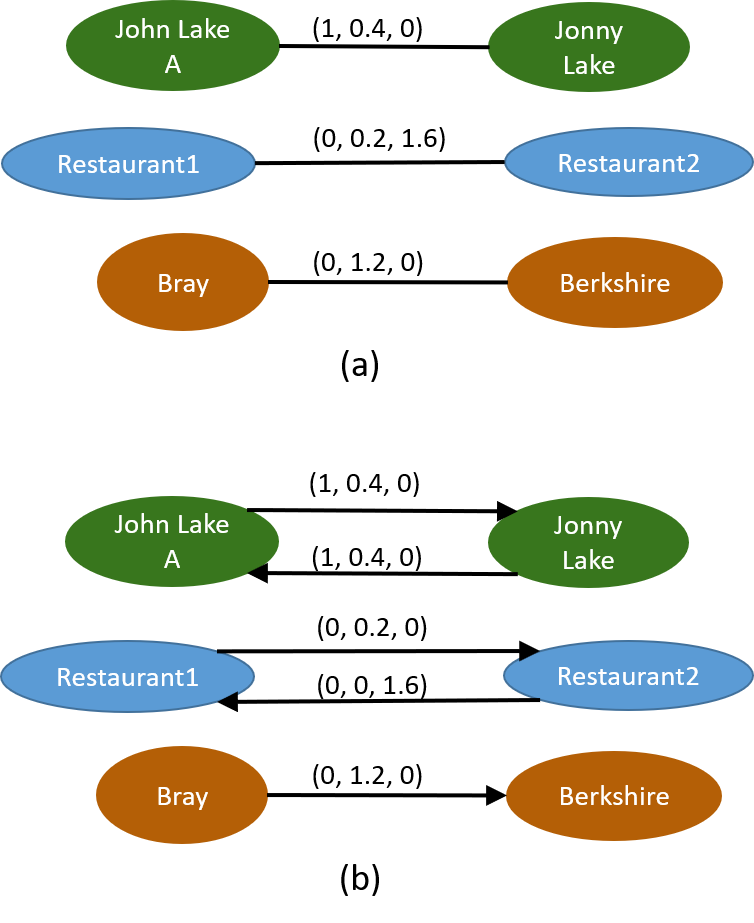}
	\caption{(a) Parts of the disjunctive blocking graph corresponding to Figure~\ref{fig:entity_graph}, and (b) the corresponding blocking graph after pruning.} 
	\vspace{-10pt}
	\label{fig:blocking_graph}
\end{figure}

\subsection{Disjunctive Blocking Graph} \label{sec:matching_blockingGraph} 
The disjunctive blocking graph $G$ is an abstraction of the disjunctive co-occurrence condition of candidate matches in blocks. Nodes represent candidates from our input entity descriptions, while edges represent pairs of candidates for which at least one of the co-occurrence conditions is `true'. Each edge is actually labeled with three weights that quantify similarity evidence on
names, tokens and neighbors of candidate matches. Specifically, the disjunctive
blocking graph of MinoanER is a graph $G = (V, E, \lambda)$,
with $\lambda$ assigning to each
edge a label ($\alpha, \beta, \gamma$), where $\alpha$ is `1' if $o_N(e_i,
e_j)$ is true and the name block in which $e_i$, $e_j$ co-occur is of 
size 2, and `0' otherwise, $\beta = valueSim(e_i,e_j)$, and $\gamma =
neighborNSim(e_i,e_j)$. More formally:

\begin{definition}
\label{def:compositeBlockingGraph}
Given a block collection $B = \bigcup_{h_{key} \in H}B^{key}$, produced by a set of indexing functions $H$, the \textbf{disjunctive blocking graph} for an entity collection $\mathcal{E}$, is a graph $G = (V, E, \lambda)$, where each node $v_i \in V$ represents a distinct description $e_i \in \mathcal{E}$, and each edge <$v_i, v_j$> $\in E$ represents a pair $e_i,e_j  \in \mathcal{E}$ for which $\mathcal{O}(e_i, e_j) = `true'$;  
$\mathcal{O}(e_i, e_j)$ is a disjunction of the atomic co-occurrence functions $o^k$ defined along with $H$, and $\lambda : E \rightarrow R^n$ is a labeling function assigning a tuple $[w^1, \ldots, w^n]$ to each edge, where $w^k$ is a weight associated with each co-occurrence function $o^k$ of $H$.
\end{definition}

Definition~\ref{def:compositeBlockingGraph} covers the cases of an entity collection $\mathcal{E}$ being composed of one, two, or more KBs. 
When matching $k$ KBs, assuming that all of them are clean, 
the disjunctive blocking graph is $k$-partite, with each of the $k$ KBs corresponding to a different independent set of nodes, i.e., there are only edges between descriptions from different KBs. 
The only information needed to match multiple KBs is to which KB every description belongs, so as to add it to the corresponding independent set. Similarly, the disjunctive blocking graph covers dirty ER, as well.

\begin{example}
Consider the graph of Figure~\ref{fig:blocking_graph}(a), which is part of the disjunctive blocking graph generated from  Figure~\ref{fig:entity_graph}. John Lake A and Jonny Lake have a common name (``J. Lake''), and there is no other entity having this name, so there is an edge connecting them with $\alpha = 1$. Bray and Berkshire have common, quite infrequent tokens in their values, so their similarity ($\beta$ in the edge connecting them) is quite high ($\beta$ = 1.2). Since Bray is a top neighbor of Restaurant1 in Figure~\ref{fig:entity_graph}, and Berkshire is a top neighbor of Restaurant2, there is also an edge with a non-zero $\gamma$ connecting Restaurant1 with Restaurant2. 
The $\gamma$ score of this edge (1.6) is the sum of the $\beta$ scores of the edges connecting Bray with Berkshire (1.2), and John Lake A with Jonny Lake (0.4).
\end{example}

\subsection{Graph Weighting and Pruning Algorithms}
Each edge in the blocking graph represents a suggested comparison between two descriptions. To reduce the number of comparisons suggested by the disjunctive blocking graph, we keep for each node the $K$ edges with the highest $\beta$ and the $K$ edges with the highest $\gamma$ weights, while \emph{pruning} edges with 
trivial weights (i.e., 
$(\alpha, \beta, \gamma$)=(0,0,0)), since they connect descriptions unlikely to match. Given that nodes $v_i$ and $v_j$ may have different top $K$ edges based on $\beta$ or $\gamma$, we consider each undirected edge in $G$ as two directed ones, with the same initial weights, and perform pruning on them.

\begin{example}
Figure~\ref{fig:blocking_graph}(b) shows the pruned version of the graph in
Figure~\ref{fig:blocking_graph}(a). Note that \textit{the blocking graph is
only a conceptual model, which we do not materialize; we
retrieve any necessary information from computationally cheap inverted indices.}
\end{example}

\input{algorithms/compositeBlockingGraphConstruction}

The process of weighting and pruning the edges of the disjunctive blocking graph is described in Algorithm~\ref{alg:compositeBlockingGraph}.
Initially, the graph contains no edges. We start adding edges by checking the name blocks 
$B_N$ (Lines~\ref{line:nameBlocking}-\ref{line:nameBlockingEnd}). For each name block $b$ that contains exactly two entities, one from each KB, we create an edge with $\alpha$=1 linking those entities
(note that in Algorithm~\ref{alg:compositeBlockingGraph}, $b^k$, $k$$\in$$\{1,2\}$, denotes the sub-block of $b$ that contains the entities from $\mathcal{E}_k$, i.e., $b^k$$\subseteq$$\mathcal{E}_k$).
Then, we compute the $\beta$ weights (Lines~\ref{line:valueBlocking}-\ref{line:valueSim}) by running a variation of Meta-blocking \cite{DBLP:journals/tkde/PapadakisKPN14}, adapted to our value similarity metric (Definition~\ref{def:valuesim}). Next, we keep for each entity, its connected nodes from the $K$ edges with the highest $\beta$ (Lines~\ref{line:firstValueLine}-\ref{line:lastValueLine}). Line~\ref{line:getInNeighbors} calls the procedure for computing the top in-neighbors of each entity, which operates as follows: first, it identifies each entity's $topNneigbors$ (Lines \ref{line:firstLineTopN}-\ref{line:lastLineTopN}) and then, it gets their reverse; for every entity $e_i$, we get the entities $topInNeighbors[i]$ that have $e_i$ as one of their $topNneighbors$ (Lines \ref{line:firstLineReverse}-\ref{line:lastLineReverse}).
This allows for estimating the $\gamma$ weights according to Definition~\ref{def:neighborsim}. To avoid re-computing the value similarities that are necessary for the $\gamma$ computations, we exploit the already computed $\beta$s. 
For each pair of entities $e_i \in \mathcal{E}_1$, $e_j \in \mathcal{E}_2$ that are connected with an edge with $\beta > 0$,
we assign to each pair of their $inNeighbors$, $(in_i, in_j)$,
a partial $\gamma$ equal to this $\beta$ (Lines~\ref{line:getInNeighbors}-\ref{line:partialGamma}). After iterating over all such entity pairs $e_i, e_j$,
we get their total neighbor similarity, i.e., 
$\gamma[i,j] = neighborNsim(e_i, e_j)$. Finally, we keep for each entity, its $K$ neighbors with the highest $\gamma$ (Lines~\ref{line:firstTopGamma}-\ref{line:lastTopGamma}).

The time complexity of Algorithm \ref{alg:compositeBlockingGraph} is dominated by the processing of value evidence, which iterates twice over all comparisons in the token blocks $B_T$. In the worst-case, this results in one computation for every pair of entities, i.e., $O(|\mathcal{E}_1| \cdot|\mathcal{E}_2|)$. In practice, though, we bound the number of computations by removing excessively large blocks that correspond to highly frequent tokens (e.g., stop-words). Following \cite{DBLP:journals/tkde/PapadakisKPN14}, this is carried out by Block Purging \cite{DBLP:journals/tkde/PapadakisIPNN13}, which ensures that the resulting blocks involve two orders of magnitude fewer comparisons than the brute-force approach, without any significant impact on recall. This complexity is higher than that of name and neighbor evidence, which are both linearly dependent on the number of input entities. The former involves a single iteration over the name blocks $B_N$, which amount to $|\mathcal{N}_1 \cap \mathcal{N}_2|$, as there is one block for every name shared by $\mathcal{E}_1$ and $\mathcal{E}_2$.  For neighbor evidence, Algorithm~\ref{alg:compositeBlockingGraph} checks all pairs of $N$ neighbors between every entity $e_i$ and its $K$ most value-similar descriptions,  performing $K \cdot N^2 \cdot (|\mathcal{E}_1|+|\mathcal{E}_2|)$ operations; the cost of estimating the top in-neighbors for each entity is lower, dominated by the ordering of all relations in $\mathcal{E}_1$ and $\mathcal{E}_2$ (i.e., $|R_{max}|\cdot log|R_{max}|$), where $|R_{max}|$ stands for the maximum number of relations in one of the KBs.

%% file: algorithms/compositeBlockingGraphConstruction.tex
\begin{algorithm}[t!]
\footnotesize
  \LinesNumbered
  \SetAlgoLined 
  \SetAlgoVlined
  \KwIn{$\mathcal{E}_1, \mathcal{E}_2$, the blocks from name and token blocking, $B_N$ and $B_T$}
  \KwOut{A disjunctive blocking graph $G$.}

\SetKwProg{myproc}{procedure}{}{}
\myproc{getCompositeBlockingGraph($\mathcal{E}_1, \mathcal{E}_2, B_N, B_T$)}{
	$V \gets \mathcal{E}_1 \cup \mathcal{E}_2$\; \label{line:initStart}
	$E \gets \emptyset$\;
	$W \gets \emptyset$ \tcp*{init. to $(0,0,0)$} \label{line:initEnd}
	\BlankLine
	\tcp{\textbf{Name Evidence}}
	\For{$b \in B_N$}{ \label{line:nameBlocking}
		\If(\tcp*[f]{only 1 comparison in $b$}){$|b^1| \cdot |b^2| = 1$}{
			$e_i$$\gets$$b^1.get(0)$, $e_j$$\gets$$b^2.get(0)$ \tcp*{entities~in~b}
      $E \gets E \cup \{<v_i, v_j>\}$\;
      $W \gets W.set(\alpha, <v_i, v_j>, 1)$\; \label{line:nameBlockingEnd}
    }
	}
	\BlankLine
	\tcp{\textbf{Value Evidence}} 
	\For{$e_i \in \mathcal{E}_1$}{ \label{line:valueBlocking}
		$\beta[] \gets \emptyset$ \tcp*{value weights wrt all $e_j \in \mathcal{E}_2$}
		\For{$b \in B_T \wedge b \cap {e_i} \neq \emptyset$}{
			\For(\tcp*[f]{ $e_j \in \mathcal{E}_2$}){$e_j \in b^2$}{ 
        	$\beta[j]$$\gets$$\beta[j]$+1/$log_2$($|b^1|$$\cdot$$|b^2|$+1) \label{line:valueSim} \tcp*{valueSim}
      }
		}
    $ValueCandidates \gets getTopCandidates(\beta[], K)$\; \label{line:firstValueLine}
    \For{$e_j \in ValueCandidates$}{
			$E \gets E \cup \{<v_i, v_j>\}$\;
      $W \gets W.set(\beta, <v_i, v_j>, \beta[j])$\; \label{line:lastValueLine}
		}
	}
	\lFor(\tcp*[f]{...do the same for $\mathcal{E}_2$}){$e_i \in \mathcal{E}_2$}{\ldots} 
	\BlankLine
	\tcp{\textbf{Neighbor Evidence}}
	$inNeighbors[] \gets getTopInNeighbors(\mathcal{E}_1,\mathcal{E}_2)$\; \label{line:getInNeighbors}
	$\gamma[][] \gets \emptyset$ \tcp*{neighbor weights wrt all $e_i, e_j \in V$ }
	\For{$e_i \in \mathcal{E}_1$}{
		\For{$e_j \in \mathcal{E}_2$, s.t. $W.get(\beta, <v_i, v_j>) > 0$}{
			\For{$in_j \in inNeighbors[j]$}{
				\For(\tcp*[f]{{\scriptsize neighborNSim}}){$in_i \in inNeighbors[i]$}{
					$\gamma[i][j] \gets \gamma[i][j]+W.get(\beta,<n_i,n_j>)$\; 
    		}
  		}
		}
	}
	\lFor(\tcp*[f]{...do the same for $\mathcal{E}_2$}){$e_i \in \mathcal{E}_2$}{\ldots}\label{line:partialGamma}
	\For {$e_i \in \mathcal{E}_1$}{ \label{line:firstTopGamma}
		$NeighborCandidates \gets getTopCandidates(\gamma[i][], K)$\;
    \For{$e_j \in NeighborCandidates$}{
    	$E \gets E \cup \{<v_i, v_j>\}$\;
      $W.set(\gamma, <v_i, v_j>, \gamma[i][j])$\; 
		}
	}
	\lFor(\tcp*[f]{...do the same for $\mathcal{E}_2$}){$e_i \in \mathcal{E}_2$}{\ldots}\label{line:lastTopGamma}
	\KwRet{$G=(V,E,W)$}\;
}
\BlankLine
\BlankLine
\myproc{getTopInNeighbors($\mathcal{E}_1, \mathcal{E}_2$)}{
	$topNeighbors[] \gets \emptyset$ \tcp*{one list for each entity} \label{line:firstLineTopN}
	$globalOrder \gets$ sort $\mathcal{E}_1$'s relations by importance\;
  \For{$e \in \mathcal{E}_1$}{
$localOrder(e)$$\gets$$relations(e).sortBy(globalOrder)$\;
    $topNrelations \gets localOrder(e).topN$\;
    \For{$(p,o) \in e$, where $p \in topNrelations$}{
			$topNeighbors[e].add(o)$\; 
    }
	}
  \lFor(\tcp*[f]{...do the same for $\mathcal{E}_2$}){$e_i \in \mathcal{E}_2$}{\ldots} \label{line:lastLineTopN}
  $topInNeighbors[]$$\gets$$\emptyset$\tcp*{{\scriptsize the reverse of topNeighbors}}\label{line:firstLineReverse}
  \For{$e \in \mathcal{E}_1 \cup \mathcal{E}_2$}{
		\For{$ne \in topNeighbors[e]$}{
			$topInNeighbors[ne].add(e)$\;\label{line:lastLineReverse}
		}
  }
  \KwRet{$topInNeighbors$}\;
}
\caption{Disjunctive Blocking Graph Construction.}\label{alg:compositeBlockingGraph}
\end{algorithm}

%% file: 4_matching.tex
\section{Non-iterative Matching Process}\label{sec:matching_matching}
Our matching method receives as input the disjunctive blocking graph $G$ and
performs four steps -- unlike most existing works, which involve a data-driven
iterative process. In every step, a matching rule is applied with the goal of extracting new matches from the edges of $G$ by analyzing their weights. The
functionality of our algorithm is outlined in Algorithm
\ref{alg:matching_algorithm}.
Next, we describe its rules in the order they are applied:

\input{algorithms/evidenceBasedMatching}

\vspace{2pt}
\noindent\textbf{Name Matching Rule (\textsf{R1}).} The matching evidence of
\textsf{R1} comes from the entity names. It assumes that \emph{two candidate
entities match, if they, and only they, have the same name $n$}. Thus, \textsf{R1}
traverses $G$ and for every edge <$v_i,v_j$> with $\alpha$ = 1, it updates the
set of matches $M$ with the corresponding descriptions (Lines
\ref{line:startH1}-\ref{line:endH1} in Alg. \ref{alg:matching_algorithm}). All
candidates matched by \textsf{R1} are not examined by the remaining rules.

\vspace{2pt}
\noindent\textbf{Value Matching Rule (\textsf{R2}).} It presumes that \emph{two entities match, if they, and only they, share a common token $t$, or, if they share many infrequent tokens.} Based on Definition~\ref{def:valuesim}, \textsf{R2} identifies pairs of descriptions with high value similarity (Lines \ref{line:startH2}-\ref{line:endH2}). To this end, it goes through every node $v_i$ of $G$ and checks whether the corresponding description stems from the smaller in size KB, for efficiency reasons (fewer checks), but has not been matched yet. In this case, it locates the adjacent node $v_j$ with the maximum $\beta$ weight (Line \ref{line:topCand}). If $\beta \geq 1$, \textsf{R2} considers the pair ($e_i, e_j$) to be a match. Matches identified by \textsf{R2} will not be considered in the sequel. 

\vspace{2pt}
\noindent\textbf{Rank Aggregation Matching Rule (\textsf{R3}).} This rule identifies further matches for candidates whose value similarity is low ($\beta < 1$), yet their neighbor similarity ($\gamma$) could be relatively high. In this respect, the order of candidates rather than their absolute similarity values are used. Its functionality appears in Lines \ref{line:startH3}-\ref{line:endH3} of Algorithm \ref{alg:matching_algorithm}. In essence, \textsf{R3} traverses all nodes of $G$ that correspond to a description that has not been matched yet. For every such node $v_i$, it retrieves two lists: the first one contains adjacent edges with a non-zero $\beta$ weight, sorted in descending order (Line \ref{line:valListH3}), while the second one includes the adjacent edges sorted in decreasing non-zero $\gamma$ weights (Line \ref{line:neiListH3}). Then, \textsf{R3} aggregates the two lists by considering the normalized ranks of their elements: assuming the size of a list is $K$, the first candidate gets the score $K/K$, the second one
$(K-1)/K$, while the last one $1/K$. Overall, each adjacent node of $v_i$ takes a score equal to the weighted summation of its normalized ranks in the two lists, as determined through the trade-off parameter $\theta \in (0,1)$  (Lines \ref{line:valRankH3} \& \ref{line:neiRankH3}): the scores of the $\beta$ list are weighted with $\theta$ and those of the $\gamma$ list with 1-$\theta$. At the end,  $v_i$ is matched with its top-1 candidate match $v_j$, i.e., the one with the highest aggregate score (Line \ref{line:endH3}). Intuitively, \textsf{R3} \emph{matches $e_i$ with $e_j$, when, based on all available evidence, there is no better candidate for $e_i$ than $e_j$.}

\vspace{2pt}
\noindent\textbf{Reciprocity Matching Rule (\textsf{R4}).} It aims to clean the
matches identified by \textsf{R1}, \textsf{R2} and \textsf{R3} by exploiting the
reciprocal edges of $G$. Given that the originally undirected graph $G$ becomes directed after pruning (as it retains the best edges per node),
a pair of nodes $v_i$ and $v_j$ are reciprocally connected when there are two
edges between them, i.e., an edge from $v_i$ to $v_j$ and an edge from $v_j$ to
$v_i$. Hence, \textsf{R4} aims to improve the precision of our algorithm based
on the rationale that two entities are unlikely to match, when one of them does
not even consider the other to be a candidate for matching. Intuitively,
\emph{two entity descriptions match, only if both of them ``agree'' that they
are likely to match.} \textsf{R4} essentially iterates over all matches detected
by the above rules and discards those missing any of the two directed edges
(Lines \ref{line:startH4}-\ref{line:endH4}), acting more like a filter for the  matches suggested by the previous rules.

Given a pruned disjunctive blocking graph, every rule can be formalized as a function that receives a pair of entities and returns true ($T$) if the entities match according to the rule's rationale, or false ($F$) otherwise, i.e., $Rn : \mathcal{E}_1\times\mathcal{E}_2 \rightarrow \{T,F\}$. In this context, we formally define the MinoanER matching process as follows: 
\begin{definition} The \textbf{non-iterative matching} of two KBs
$\mathcal{E}_1$, $\mathcal{E}_2$, denoted by the Boolean matrix $M(  \mathcal{E}_1, \mathcal{E}_2)$, is defined as a filtering problem of the pruned disjunctive blocking graph $G$:
$M(e_i,e_j) = (\textsf{R1}(e_i,e_j) \vee \textsf{R2}(e_i,e_j) \vee \textsf{R3}(e_i,e_j)) \wedge \textsf{R4}(e_i,e_j)$.
\end{definition} 

The time complexity of Algorithm~\ref{alg:matching_algorithm} is dominated by
the size of the pruned blocking graph $G$ it receives as input, since
\textsf{R1}, \textsf{R2} and \textsf{R3} essentially go through all directed
edges in $G$ (in practice, though, \textsf{R1} reduces the edges considered by
\textsf{R2} and \textsf{R3}, and so does \textsf{R2} for \textsf{R3}). In the
worst case, $G$ contains $2K$ directed edges for every description in
$\mathcal{E}_1 \cup \mathcal{E}_2$, i.e., $|V|_{max}=2\cdot K \cdot (|\mathcal{E}_1|+|\mathcal{E}_2|)$. Thus, the overall complexity is linear with
respect to the number of input descriptions, i.e.,
$O($$|\mathcal{E}_1|$$+$$|\mathcal{E}_2|$$)$, yielding high scalability.

\begin{figure}[t]
	\center \includegraphics[width=0.9\columnwidth]{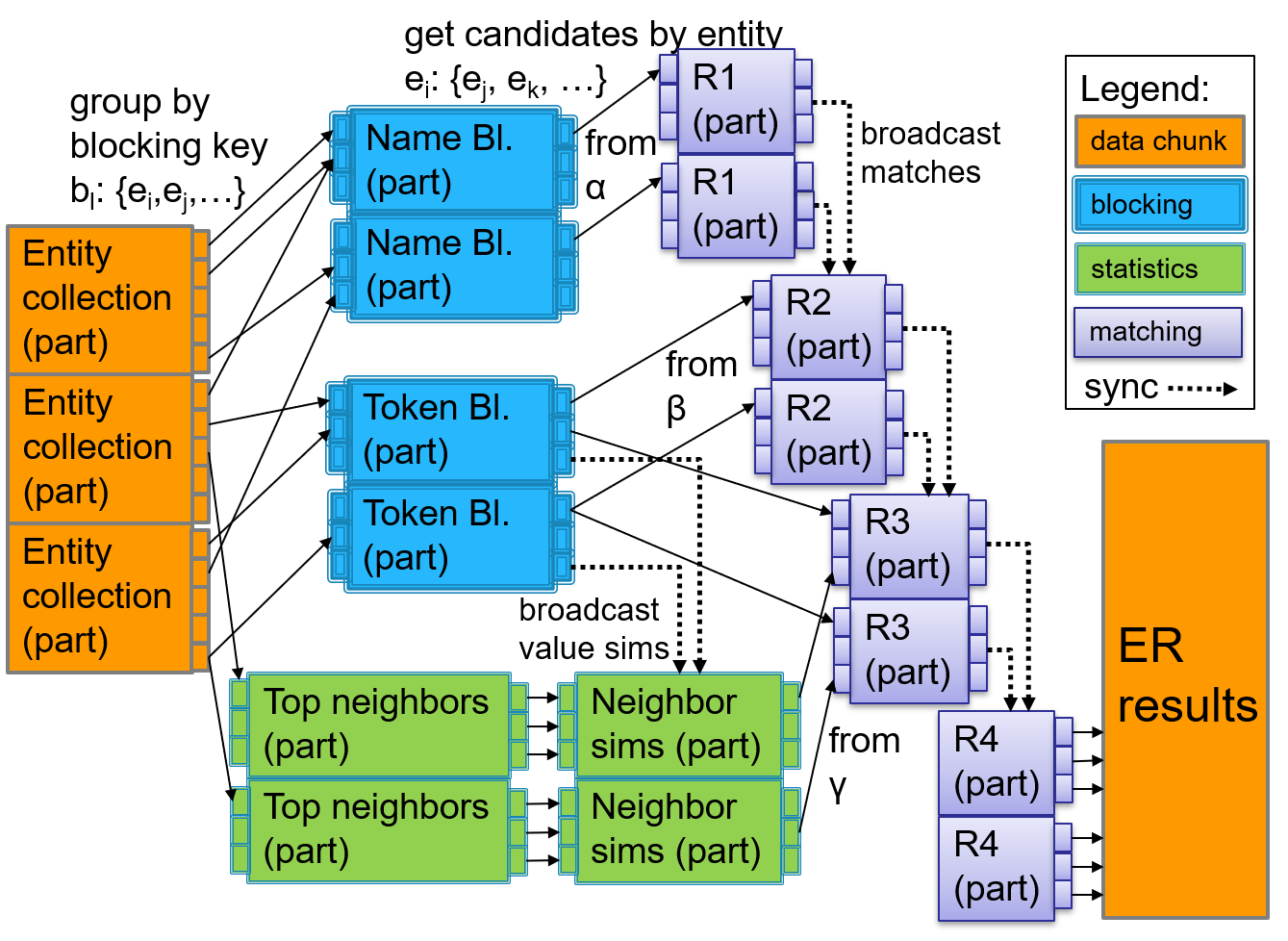}
	\caption{The architecture of MinoanER in Spark.}
	\vspace{-10pt}
	\label{fig:architecture}
    
\end{figure}

\subsection{Implementation in Spark}
\label{ssec:matching_implementation}

Figure~\ref{fig:architecture} shows the architecture of MinoanER implementation in Spark. 
Each process is executed in parallel for different chunks of input, in different Spark workers. 
Each dashed edge represents a synchronization point, at which the process has to wait for results produced by different data chunks (and different Spark workers). 

In more detail, Algorithm~\ref{alg:compositeBlockingGraph} is adapted to Spark by applying name blocking simultaneously with token blocking and the extraction of top neighbors per entity. Name blocking and token blocking produce the sets of blocks $B_N$ and $B_T$, respectively, which are part of the algorithm's input.
The processing of those blocks
in order to estimate the $\alpha$ and $\beta$ weights
(Lines 5-9 for $B_N$ and Lines 10-18 for $B_T$) takes place during the construction of the blocks. The extraction of top neighbors per entity (Line 20) runs in parallel to these two processes and its output, along with the $\beta$ weights,
is given to the last part of the graph construction, which computes the $\gamma$ weights for all entity pairs with neighbors co-occuring in at least one block (Lines 21-33). 

To minimize the overall run-time, Algorithm~\ref{alg:matching_algorithm} is adapted to Spark as follows:
\textsf{R1} (Lines 2-4)
is executed in parallel with name blocking and the matches it discovers 
are broadcasted to be excluded from subsequent rules. \textsf{R2} (Lines 5-9) 
runs after both \textsf{R1} and token blocking have finished, while \textsf{R3} (Lines 10-23)
runs after both \textsf{R2} and the computation of neighbor similarities have been completed, skipping the already identified (and broadcasted) matches.
\textsf{R4} (Lines 24-26)
runs in the end, providing the final, filtered set of matches. 
Note that during the execution of every rule, each Spark worker contains only the partial information of the disjunctive blocking graph that is necessary to find the match of a specific node (i.e., the corresponding lists of candidates based on names, values, or neighbors).

%% file: algorithms/evidenceBasedMatching.tex
\begin{algorithm}
\footnotesize
  \LinesNumbered
  \SetAlgoLined 
  \SetAlgoVlined
  \KwIn{$\mathcal{E}_1, \mathcal{E}_2$, The pruned, directed disjunctive blocking graph $G$.}
  \KwOut{A set of matches $M$.}

$M \gets \emptyset$\tcp*{The set of matches}
\BlankLine
\tcp{\textbf{Name Matching Value (\textsf{R1})}}
\For {$<v_i,v_j> \in G.E$}{ \label{line:startH1}
	\If {$G.W.get(\alpha, <v_i,v_j>) = 1$}{
		$M \gets M \cup (e_i, e_j)$\; \label{line:endH1}
	}
}
\BlankLine
\tcp{\textbf{Value Matching Value (\textsf{R2})}}
\For{$v_i \in G.V$}{ \label{line:startH2}
	\If(\tcp*[f]{check the smallest KB for efficiency}){$e_i \in \mathcal{E}_1 \setminus M$}{
		$v_j \gets argmax_{v_k \in G.V}G.W.get(\beta, <v_i,v_k>)$ \tcp*{top candidate} \label{line:topCand}
    \If{$G.W.get(\beta, <v_i,v_j>) \geq 1$}{
			$M \gets M \cup (e_i, e_j)$\; \label{line:endH2}
    }
  }
} 
\BlankLine
\tcp{\textbf{Rank Aggregation Matching Value (\textsf{R3})}}
\For{$v_i \in G.V$}{\label{line:startH3}
	\If{$e_i \in \mathcal{E}_1 \cup \mathcal{E}_2 \setminus M$}{
		$agg[] \gets \emptyset$\tcp*{Aggregate scores, init. zeros}
    $valCands \gets G.valCand(e_i)$ \tcp*{nodes linked to $e_i$ in decr.  $\beta$} \label{line:valListH3}
    $rank \gets |valCands|$\;
    \For{$e_j \in valCands$}{
			$agg[e_i].update(e_j, \theta \cdot rank/|valCands|)$\; \label{line:valRankH3}
			$rank \gets rank - 1$\;
    }
    $ngbCands \gets G.ngbCand(e_i)$ \tcp*{nodes linked to $e_i$ in decr. $\gamma$} \label{line:neiListH3}
    $rank \gets |ngbCands|$\;
    \For {$e_j \in ngbCands$}{
			$agg[e_i].update(e_j, (1-\theta) \cdot rank/|ngbCands|)$\; \label{line:neiRankH3}
			$rank \gets rank - 1$\;
    }
	}
  $M \gets M \cup (e_i, getTopCandidate(agg[e_i]))$\; \label{line:endH3}
}
\BlankLine
\tcp{\textbf{Reciprocity Matching Value (\textsf{R4})}}
\For {$(e_i,e_j) \in M$}{ \label{line:startH4}
	\If{$<v_i,v_j> \notin G.E$ $\vee <v_j,v_i> \notin G.E$}{
		$M \gets M \setminus (e_i, e_j)$\; \label{line:endH4}
  }
}
\KwRet{$M$}\; \label{line:end}
\caption{Non-iterative Matching.}\label{alg:matching_algorithm}
\end{algorithm}

%% file: 5_relatedWork.tex
\section{Related Work}\label{sec:matching_relatedWork}
To the best of our knowledge, there is currently no other Web-scale ER framework that is fully automated, non-iterative, schema-agnostic and 
massively parallel, at the same time. 
For example, WInte.r~\cite{DBLP:conf/semweb/LehmbergBB17} is a framework that performs multi-type ER, also incorporating the steps of blocking, schema-level mapping and data fusion. However, it is implemented in a sequential fashion and its solution relies on a specific level of structuredness, i.e., on a schema followed by the instances to be matched. 
Dedoop~\cite{DBLP:journals/pvldb/KolbTR12} is a highly scalable ER framework that comprises the steps of blocking and supervised matching. However, it is the user that is responsible for selecting one of the available blocking and learning methods and for fine-tuning their internal parameters.
This approach is also targeting datasets with a predefined schema. Dedupe~\cite{Dedupe} is a scalable open-source Python library (and a commercial tool built on this library) for ER; however, it is not fully automated, as it performs active learning, relying on human experts to decide for a first few difficult matching decisions. Finally, we consider progressive ER (e.g.,~\cite{Altowim:2018:PAP:3178546.3154410}) orthogonal to our approach, as it aims to retrieve as many matches as possible as early as possible.

In this context, we compare MinoanER independently to state-of-the-art matching and blocking approaches for Web data. 

\textbf{Entity Matching.} Two types of similarity measures are commonly used for entity matching \cite{DBLP:conf/kdd/Lacoste-JulienPDKGG13,DBLP:journals/jcst/ShaoHLWCX16}.
(i) \emph{Value-based similarities} (e.g., Jaccard, Dice) usually assess the
similarity of two descriptions based on the values of specific attributes.  Our
value similarity is a variation of ARCS, a Meta-blocking weighting
scheme~\cite{DBLP:journals/is/Efthymiou0PSP17}, which disregards any schema
information and considers each entity description as a bag of words. Compared to ARCS,
though, 
we focus more on the \emph{number than the frequency of
common tokens between two descriptions}. (ii) \emph{Relational similarity} measures additionally consider neighbor similarity by exploiting the value similarity of 
(some of) the entities' neighbors. 


Based on the nature of the matching decision, 
ER can be characterized as \emph{pairwise} or \emph{collective}. 
The former relies exclusively on 
the value similarity of descriptions to decide if they match (e.g.,~\cite{DBLP:journals/pvldb/KolbTR12}), while the latter
iteratively updates the matching decision for entities by dynamically assessing the similarity of their neighbors (e.g.,~\cite{DBLP:journals/tkdd/BhattacharyaG07}). Typically, the starting point for this similarity propagation is a set of seed matches identified by a value-based blocking. 


For example, SiGMa~\cite{DBLP:conf/kdd/Lacoste-JulienPDKGG13} starts with seed matches having identical entity names. Then, it propagates the matching decisions on the `compatible' neighbors,
which are linked with pre-aligned relations.
For every new matched pair, the similarities of the neighbors are recomputed and their position in the priority queue is updated. 
LINDA~\cite{DBLP:conf/cikm/BohmMNW12} differs by considering as compatible
neighbors those connected with relations having similar names (i.e., small edit distance). However, this requirement rarely holds in the extreme schema heterogeneity of Web data.
RiMOM-IM~\cite{DBLP:journals/jcst/ShaoHLWCX16,DBLP:journals/tkde/LiTLL09} 
is a similar approach, 
introducing the following heuristic: if two matched descriptions $e_1, e_1'$ are connected via aligned relations $r$ and $r'$ and all their entity neighbors via $r$ and $r'$, except $e_2$ and $e_2'$, have been matched, then $e_2$ and $e_2'$ are also considered matches. 

All these methods employ \textsf{Unique Mapping Clustering} for detecting matches: they place all pairs into a priority queue, in decreasing
similarity. At each iteration, the top pair is considered a match, if none of its entities has been already matched. The process ends when the top pair has a similarity lower than $t$.

MinoanER employs \textsf{Unique Mapping Clustering}, too. Yet, 
it differs from SiGMa, LINDA and RiMOM-IM in five ways:
(i) the matching process iterates over the disjunctive blocking graph, instead of the initial KBs. (ii) MinoanER employs statistics to automatically discover distinctive entity names and important relations. (iii) MinoanER exploits different sources of matching evidence (values, names and neighbors) to statically identify candidate matches already from the step of blocking. (iv) MinoanER does not aggregate different similarities in one similarity score; instead, it uses a disjunction of the different evidence it considers.
(v) MinoanER is a static collective ER approach, in which all sources of similarity are assessed only once per candidate pair. By considering a composite blocking not only on the value but also on the neighbors similarity, we discover in a non-iterative way most of the matches returned by the data-driven convergence of existing systems, or even more (see Section~\ref{sec:matching_experiments}).

PARIS~\cite{DBLP:journals/pvldb/SuchanekAS11} uses a probabilistic model to identify matches, based on previous matches and the functional nature of entity relations. A relation is considered \textit{functional} if, for a source entity, there is only one destination entity. 
If $r(x,y)$ is a function in one KB and $r(x,y')$ a function in another KB, then $y$ and $y'$ are considered matches. The functionality of a relation and the alignment of relations along with previous matching decisions determine the decisions in subsequent iterations. 
\emph{Unlike MinoanER, PARIS cannot deal with structural heterogeneity,  while it targets both ontology and instance matching}.

Finally, \cite{DBLP:journals/pvldb/RastogiDG11} parallelizes the collective ER approach of~\cite{DBLP:journals/tkdd/BhattacharyaG07}, relying on a black-box matching and exploits a set of heuristic rules for structured entities. It essentially runs multiple instances of the matching algorithm in subsets of the input entities (similar to blocks), also keeping information for all the entity neighbors, needed for similarity propagation. Since some rules may require the results of multiple blocks, an iterative message-passing framework is employed. \emph{Rather than a block-level synchronization, the MinoanER parallel computations in Spark require synchronization only across the 4 
generic matching rules} 
(see Figure~\ref{fig:architecture}).

Regarding the matching rules, the ones employed by MinoanER based on values and names are similar to rules that have already been employed in the literature individually (e.g., in \cite{DBLP:conf/kdd/Lacoste-JulienPDKGG13,DBLP:journals/tkde/LiTLL09,DBLP:journals/jcst/ShaoHLWCX16}). In this work, we use a combination of those rules for the first time, also introducing a novel rank aggregation rule to incorporate value and neighbor matching evidence. Finally, the idea of reciprocity has been applied to enhance the results of Meta-blocking~\cite{DBLP:journals/bdr/PapadakisPPK16}, but was never used in matching.

\textbf{Blocking.} 
Blocking techniques for relational
databases~\cite{DBLP:books/daglib/0030287} rely on blocking keys defined at the schema-level. 
For example, the \textit{Sorted Neighborhood} approach
orders entity descriptions according to a sorting criterion and performs blocking based on it; it is expected that matching descriptions will be neighbors after the sorting, so neighbor descriptions constitute candidate matches~\cite{DBLP:conf/sigmod/HernandezS95}. Initially, entity descriptions are lexicographically ordered based on their blocking keys. Then, a window, resembling a block, of fixed length slides over the ordered descriptions, each time comparing only the contents of the window. An adaptive variation of the sorted neighborhood method is to dynamically decide on the size of the window~\cite{DBLP:dblp_conf/jcdl/YanLKG07}. In this case, adjacent blocking keys in the sorted descriptions that are significantly different from each other, are used as boundary pairs, marking the positions where one window ends and the next one starts. Hence, this variation creates non-overlapping blocks. In a similar line of work, the sorted blocks method \cite{DBLP:conf/nss/DraisbachN11} allows setting the size of the window, as well as the degree of desired overlap.

Another recent schema-based blocking method 
uses Maximal Frequent Itemsets (MFI) as blocking keys \cite{DBLP:journals/is/KenigG13} -- an itemset can be a set of tokens. Abstractly, each MFI of a specific attribute in the schema of a description defines a block, and descriptions containing the tokens of an MFI for this attribute are placed in a common block. Using frequent itemsets to construct blocks may significantly reduce the number of candidates for matching pairs. However, since many matching descriptions share few, or even no common tokens, further requiring that those tokens are parts of frequent itemsets is too restrictive.
The same applies to the requirement for a-priori schema knowledge and alignment, thus
resulting in many missed matches in the Web of Data. 

Although blocking has been extensively studied for tabular data, the proposed approaches cannot be used for the Web of Data, since their blocking keys rely on the existence of a global schema.
However, 
the use of schema-based blocking keys is inapplicable to the Web of Data, due to its extreme schema heterogeneity \cite{DBLP:conf/pods/GolshanHMT17}:
entity descriptions do not follow a fixed schema, 
as
even a single description typically uses attributes defined in multiple LOD vocabularies. 
In this context, schema-agnostic blocking methods are needed instead. 
Yet, the schema-agnostic functionality of most blocking methods
requires extensive fine-tuning to achieve high effectiveness \cite{DBLP:journals/pvldb/0001SGP16}. 
The only exception is token
blocking, which is completely parameter-free \cite{DBLP:journals/tkde/PapadakisIPNN13}.
Another advantage of token blocking 
is that it 
allows for computing value
similarity from its blocks, as they contain entities with \textit{identical} blocking
keys -- unlike other methods like Dedoop~\cite{DBLP:journals/pvldb/KolbTR12} and Sorted Neighborhood \cite{DBLP:conf/sigmod/HernandezS95},
whose blocks contain entities with \textit{similar} keys.


SiGMa~\cite{DBLP:conf/kdd/Lacoste-JulienPDKGG13} considers descriptions with at least two common tokens as candidate matches, which is 
more precise than our token blocking, but misses more matches. The missed
matches will be considered in subsequent iterations, if their neighbor
similarity is strong, whereas \emph{MinoanER identifies such matches from the step of blocking}.
RiMOM-IM~\cite{DBLP:journals/jcst/ShaoHLWCX16} computes the tokens' TF-IDF weights, takes the top-5 tokens of each entity, and constructs a block for each one, along with the attribute this value appears. \emph{Compared to the full automation of MinoanER, this method requires attribute alignment}. 
\cite{mcneill2012dynamic} iteratively splits large blocks into smaller ones by
adding attributes to the blocking key. This leads to a prohibitive technique
for voluminous KBs of high Variety.

Disjunctive blocking schemes have been proposed for KBs of
high~\cite{DBLP:journals/corr/Kejriwal16} and
low~\cite{DBLP:conf/icdm/BilenkoKM06} levels of schema heterogeneity. 
Both methods, though, are of limited applicability, as they require labelled instances for their supervised learning.
\emph{In contrast, MinoanER copes with the Volume and Variety of the
Web of Data, 
through an unsupervised, schema-agnostic, disjunctive
blocking}.

Finally, LSH blocking techniques (e.g.,~\cite{DBLP:conf/edbt/MalhotraAS14}) hash descriptions multiple times, using a family of hash functions, so that similar descriptions are more likely to be placed into the same bucket than dissimilar ones. 
This requires tuning a similarity threshold between entity pairs, above which they are considered candidate matches. \emph{This tuning is non-trivial, especially for descriptions from different domains, while its effectiveness is limited for nearly similar entities} (see Figure~\ref{fig:similarityDistributions}). 

%% file: 6_experiments.tex
\section{Experimental Evaluation}\label{sec:matching_experiments}

\input{tables/datasets.tex}

In this section, we thoroughly compare MinoanER to state-of-the-art tools and a heavily fine-tuned baseline method.

\vspace{2pt}
\noindent
\textbf{Experimental Setup. }All experiments were performed on top of Apache Spark v2.1.0 and Java 8, on a cluster of 4 Ubuntu 16.04.2 LTS servers. Each server has 236GB RAM and 36 Intel(R) Xeon(R) E5-2630 v4 @2.20GHz CPU cores. 

\begin{figure*}[t]\centering
    \includegraphics[width=0.38\linewidth]{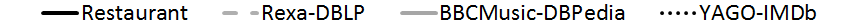}\\
	\includegraphics[width=0.245\linewidth]{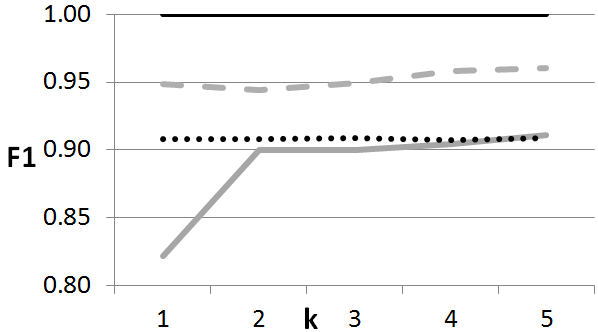}
	\includegraphics[width=0.245\linewidth]{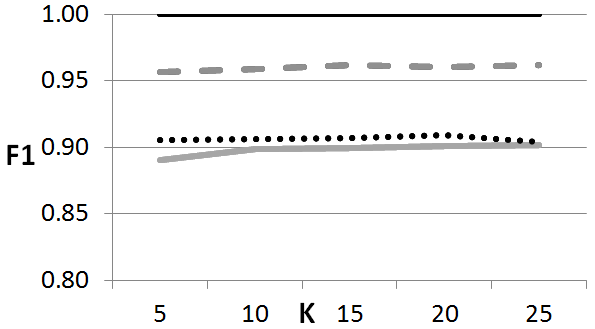}
	\includegraphics[width=0.245\linewidth]{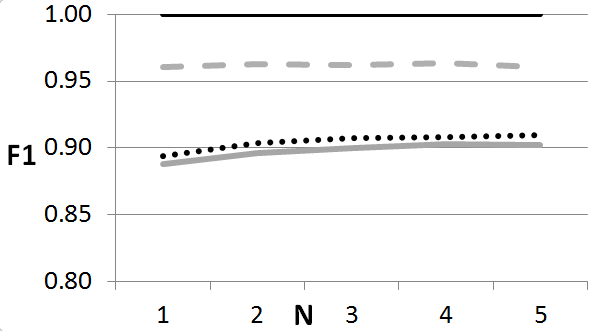}
	\includegraphics[width=0.245\linewidth]{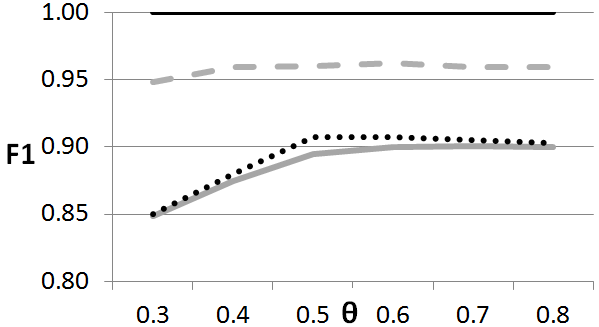}
	\caption{Sensitivity analysis of the four configuration parameters of our MinoanER.}
	\label{fig:sensAnalysis}
\vspace{-10pt}
\end{figure*} 

\vspace{2pt}
\noindent
\textbf{Datasets.} 
We use four established benchmark datasets with entities from real KBs. Their technical characteristics appear in Table~\ref{tab:matching_datasets}.
All KBs contain relations between the described entities. 

\emph{Restaurant}\footnote{http://oaei.ontologymatching.org/2010/im/} contains descriptions of restaurants and their addresses from two different KBs. 
It is the smallest dataset in terms of the number of entities, triples, entity types\footnote{Extracted using the attribute \url{w3.org/1999/02/22-rdf-syntax-ns\#type}.},
as well as the one using the smallest number of vocabularies.
We use it for two reasons: \emph{(i)} it is a popular benchmark, created by the Ontology Alignment Evaluation Initiative, and \emph{(ii)} it offers a good example of a dataset with very high value and neighbor similarity between matches (Figure~\ref{fig:similarityDistributions}), involving the easiest pair of KBs to resolve.

\emph{Rexa-DBLP}\footnote{oaei.ontologymatching.org/2009/instances/} contains descriptions of publications and their authors. The ground truth contains matches between both publications and authors. This dataset contains strongly similar matches in terms of values and neighbors (Figure~\ref{fig:similarityDistributions}). Although it is relatively easy to resolve, Table~\ref{tab:matching_datasets} shows that it exhibits the greatest difference with respect to the size of the KBs to be matched (DBLP is 2 orders of magnitude bigger than Rexa in terms of descriptions, and 3 orders of magnitude in terms of triples). 

\emph{BBCmusic-DBpedia}~\cite{DBLP:conf/bigdataconf/EfthymiouSC15} contains
descriptions of musicians, bands and their birthplaces, from BBCmusic and the
BTC2012 version of DBpedia\footnote{\url{datahub.io/dataset/bbc-music},
\url{km.aifb.kit.edu/projects/btc-2012/}}. In our experiments, we consider only
entities appearing in the ground truth, as well as their immediate in- and
out-neighbors. The most challenging characteristic of this dataset is the high
heterogeneity between its two KBs
in terms of both schema and values:
DBpedia contains $\sim$11,000 different attributes, $\sim$60,000 entity types,
953 relations, the highest number of different vocabularies (6), while using on
average 4 times more tokens than BBCmusic to describe an entity.
The latter feature means that all normalized, set-based
similarity measures like Jaccard fail to identify such matches, since a big difference in
the token set sizes yields low similarity values (see Figure~\ref{fig:similarityDistributions}). A thorough investigation has shown
that in the median case, an entity description in this dataset contains only 2
words in its values that are used by both
KBs~\cite{DBLP:conf/bigdataconf/EfthymiouSC15}.

\emph{YAGO-IMDb} \cite{DBLP:journals/pvldb/SuchanekAS11} contains descriptions of movie-related entities (e.g., actors, directors, movies) from YAGO\footnote{\url{www.yago-knowledge.org}} and IMDb\footnote{\url{www.imdb.com}}. Figure~\ref{fig:similarityDistributions} shows that a large number of matches in this dataset has low value similarity, while a significant number has high neighbor similarity. Moreover, this is the biggest dataset in terms of entities and triples, challenging the scalability of ER tools, while it is the most balanced pair of KBs with respect to their relative size. 

\input{tables/blockStats.tex}

\vspace{2pt}
\noindent
\textbf{Baselines.}
In our experiments, we compare MinoanER against four state-of-the-art methods: SiGMa, PARIS, LINDA and RiMOM. PARIS is openly available, so we ran its original implementation. 
For the remaining tools, we report their performance from the original publications\footnote{RiMOM-IM~\cite{DBLP:journals/jcst/ShaoHLWCX16} is openly available, but no execution instructions were made available to us.}.
We also consider \textsf{BSL}, a custom baseline method that receives as input the disjunctive blocking graph $G$, before pruning, and compares every pair of descriptions connected by an edge in $G$. The resulting similarities are then processed by \textsf{Unique Mapping Clustering}. Unlike MinoanER, though, \textsf{BSL} disregards all evidence from entity neighbors, relying exclusively on value similarity. Yet, it optimizes its performance through a series of well-established string matching methods that undergo extensive fine-tuning on the basis of the ground-truth.

In more detail, we consider numerous configurations for the four parameters of
\textsf{BSL} in order to 
maximize its F1: \emph{(i)} The schema-agnostic representation of the values in every entity. \textsf{BSL} uses token $n$-grams for this purpose, with $n \in \{1, 2, 3\}$, thus representing every resource by the token uni-/bi-/tri-grams that appear in its values.
\emph{(ii)} The weighting scheme that assesses the importance of every token. We consider TF and TF-IDF weights. \emph{(iii)} The similarity measure, for which we consider the following 
well-established
similarities: Cosine,
Jaccard,
Generalized Jaccard
and SiGMa (which applies exclusively to TF-IDF weights \cite{DBLP:conf/kdd/Lacoste-JulienPDKGG13}).
All measures are normalized to $[0,1]$.
\emph{(iv)} The similarity threshold that prunes the entity pairs processed by \textsf{Unique Mapping Clustering}. We use all thresholds in $[0,1)$ with a step of 0.05. In total, we consider 420 different configurations,
reporting the 
highest F1.

\subsection{Effectiveness Evaluation}
\label{ssec:results}

\textbf{Blocks Performance.} First, we examine the performance of the blocks used by MinoanER (and \textsf{BSL}). Their statistics appear in Table \ref{tab:blocks}.
We observe that the number of comparisons in token blocks ($||B_T||$) is at least 1 order of magnitude larger than those of name blocks ($||B_N||$), even if the latter may involve more blocks ($|B_N|{>}|B_T|$ over YAGO-IMDb). In fact, 
$||B_N||$ seems to depend linearly on the number of
input descriptions, whereas $||B_T||$ seems to depend
quadratically on that number. Nevertheless, the overall comparisons in $B_T\cup
B_N$ are at least 2 orders of magnitude lower than the Cartesian product
$|E_1|\cdot|E_2|$, even though recall is consistently higher than 99\%. On the flip side,
both precision and F-Measure (F1) remain rather low.

\vspace{2pt}
\noindent
\textbf{Parameter Configuration.}
Next, we investigate the robustness of our method with respect to its internal configuration. To this end, we perform a \textit{sensitivity analysis}, using the following meaningful values for the four parameters of MinoanER: $k \in \{1,2,3,4,5\}$ (the number of most distinct predicates per KB whose values serve as names), $K \in \{5, 10, 15, 2, 25\}$ (the number of candidate matches per entity from values and neighbors), $N \in \{1,2,3,4,5\}$ (the number of the most important relations per entity), and $\theta \in \{0.3, 0.4, 0.5, 0.6, 0.7, 0.8\}$ (the trade-off between value- vs neighbor-based candidates). Preliminary experiments demonstrated that the configuration $(k, K, N, \theta)=(2, 15, 3, 0.6)$ yields a performance very close to the average one. Therefore, we use these parameter values as the default ones in our sensitivity analysis.

In more detail, we sequentially vary the values of one parameter, keeping the others fixed to their default value, so as to examine its effect on the overall F1 of MinoanER. The results appear in the diagrams of Figure \ref{fig:sensAnalysis}. We observe that MinoanER is quite robust in most cases, as small changes in a parameter value typically lead to an insignificant change in F1. This should be attributed to the composite functionality of MinoanER and its four matching rules,
in particular: even if one rule is misconfigured, the other rules make up for the lost matches. There are only two exceptions:

\emph{(i)} There is a large increase in F1 over BBCmusic-DBpedia when $k$ increases from 1 to 2. The former value selects completely different predicates as names for the two KBs, due to the schema heterogeneity of DBpedia, thus eliminating the contribution of the name matching rule to F1. This is ameliorated for $k$=2. 

\emph{(ii)} F1 is significantly lower over BBCmusic-DBpedia and YAGO-IMDb for $\theta < 0.5$. This should be expected, since Figure \ref{fig:similarityDistributions} demonstrates that both datasets are dominated by nearly-similar matches, with the value similarity providing insufficient evidence for detecting them. Hence, $\theta$ should promote neighbor-similarity at least to the same level as the value-similarity (i.e., $\theta \geq 0.5$).

\input{tables/baselines.tex}

As a result, next, we can exclusively consider the configuration $(k, K, N, \theta)=(2, 15, 3, 0.6)$ for MinoanER. This is the suggested global configuration that works well over all datasets, but parameter tuning per individual dataset may yield better results.


\vspace{2pt}
\noindent
\textbf{Comparison with Baselines.} 
Table~\ref{tab:matching_results} shows that MinoanER offers competitive
performance when matching KBs with few attributes and entity types, despite
requiring no domain-specific input. Specifically, it achieves $100\%$ F1 in Restaurant, which is $3\%$ higher than SiGMa, $9\%$
higher than PARIS, and $\sim$$20\%$ higher than LINDA and RiMOM. \textsf{BSL} also achieves perfect F1, due to the strongly similar matches
(see Figure~\ref{fig:similarityDistributions}). 
In Rexa-DBLP, MinoanER also
outperforms all existing ER methods. It is $2\%$ better than SiGMa in F1,
$4.6\%$ better than PARIS, $20\%$ better than RiMOM, and 6\% better than
\textsf{BSL}. 
In YAGO-IMDb, MinoanER achieves similar
performance to SiGMa ($91\%$ F1), with more identified matches ($91\%$ vs
$85\%$), but lower precision ($91\%$ vs $98\%$). Compared to PARIS, its F1 is
$1\%$ lower, due to $3\%$ lower precision, despite the $1\%$ better recall. 
The high structural similarity between the two KBs make this dataset a good use case for PARIS. 
\textsf{BSL} exhibits the worst performance, due to the very low value
similarity of matches in this KB.
Most importantly, MinoanER achieves the
best performance by far over 
the highly heterogeneous KBs of BBCmusic-DBpedia. PARIS struggles to identify the matches,
with \textsf{BSL} performing significantly better, but still poorly in absolute
numbers. In contrast, MinoanER succeeds in identifying $89\%$ of matches with
$91\%$ precision, achieving a $90\%$ F1.

Comparing the performance of MinoanER 
to that of its input blocks,
precision raises by several orders of magnitude at the cost of slightly lower recall. 
The lower recall is caused by missed matches close to the lower left corner of Figure~\ref{fig:similarityDistributions}, i.e., with very low value and neighbor similarities. This explains why the impact on recall is larger for BBCmusic-DBpedia and YAGO-IMDb. 

\vspace{2pt}
\noindent
\textbf{Evaluation of Matching Rules.}
\label{sssec:heuristics}
Table~\ref{tab:heuristics} summarizes the performance of each matching rule in Algorithm~\ref{alg:matching_algorithm}, when executed alone, as well as the overall contribution of neighbor similarity evidence.

\input{tables/matchingRules.tex}
\begin{figure*}[t]\centering
	\includegraphics[width=0.245\linewidth]{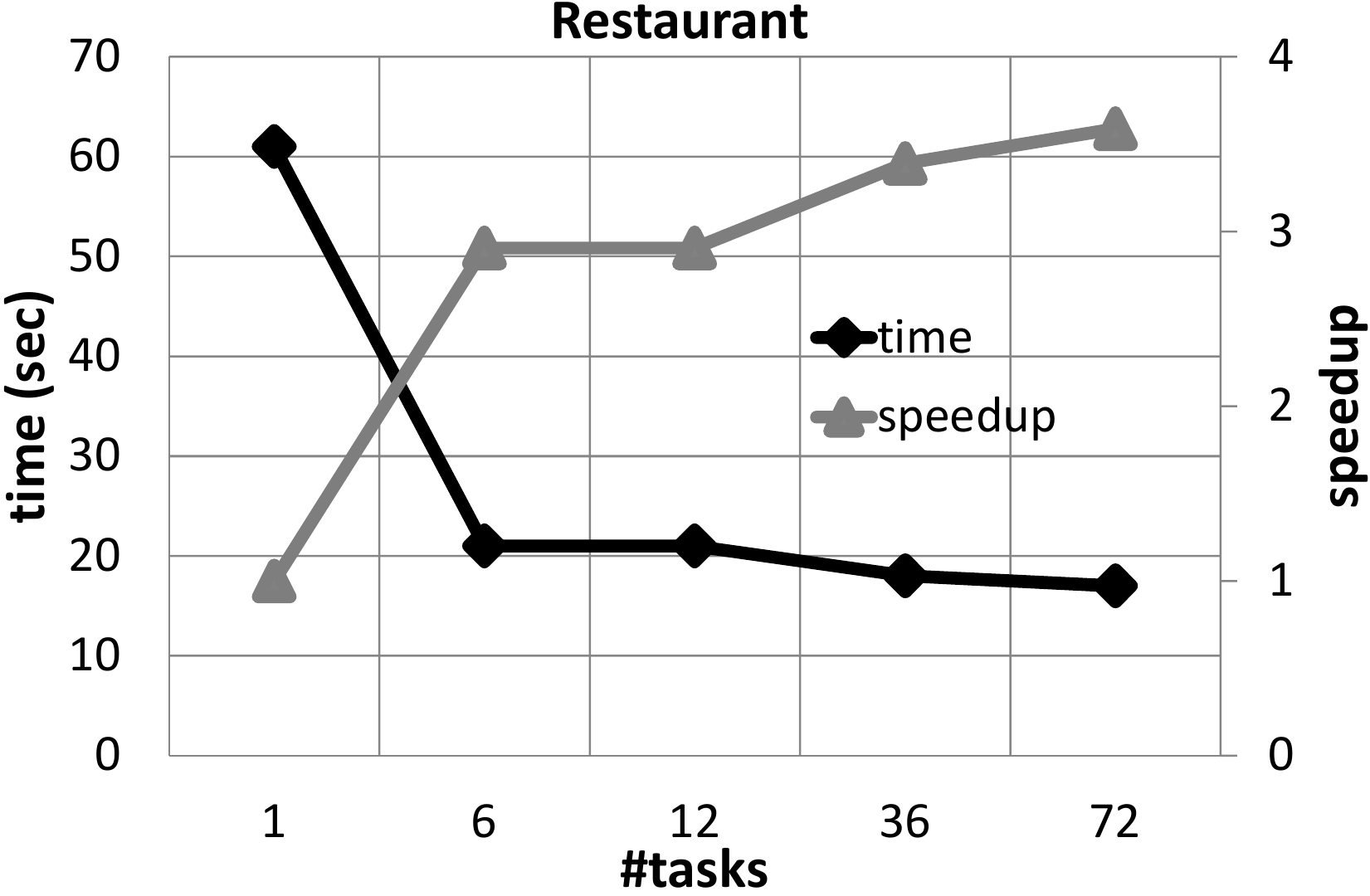}
	\includegraphics[width=0.245\linewidth]{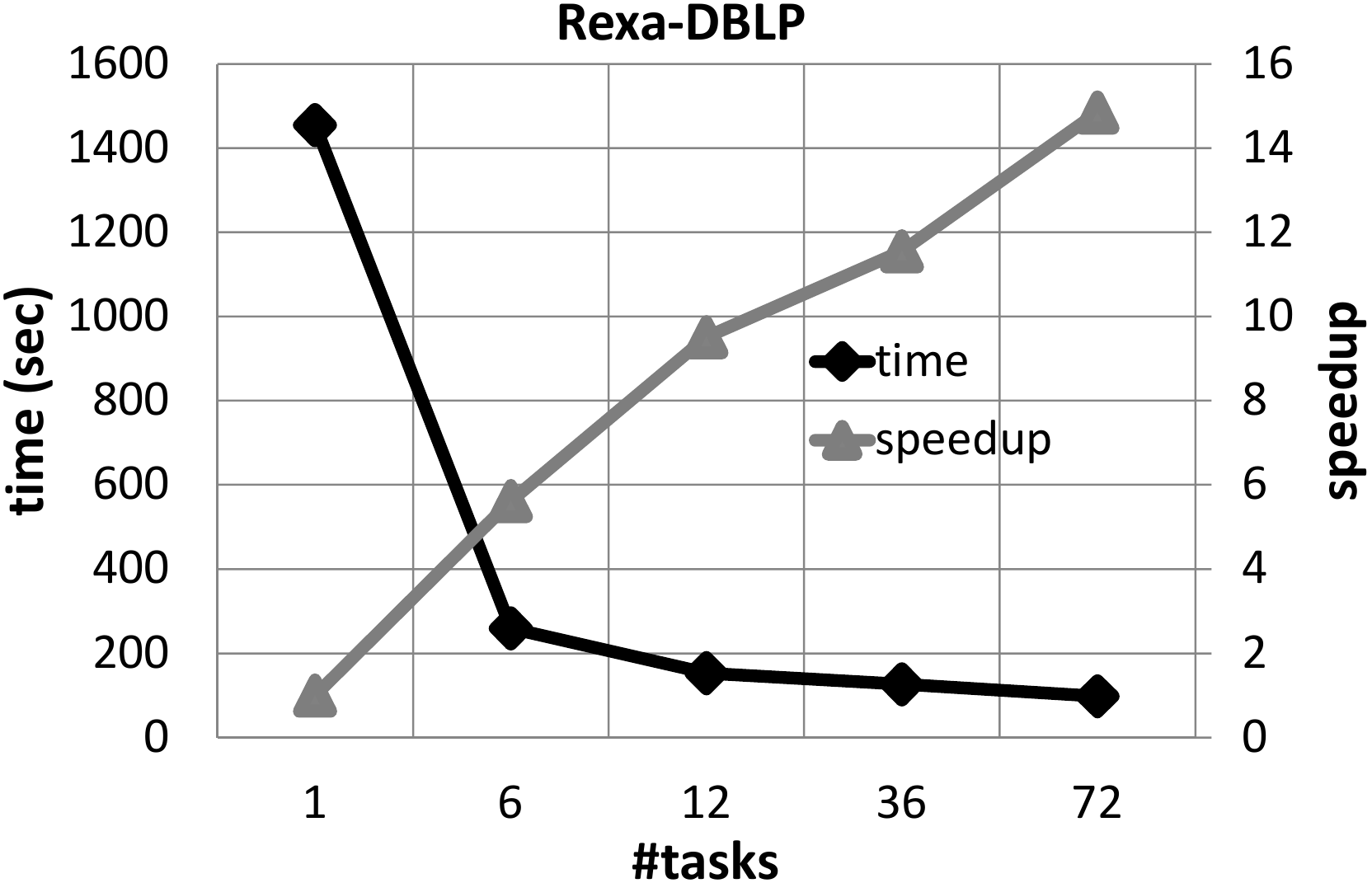}
	\includegraphics[width=0.245\linewidth]{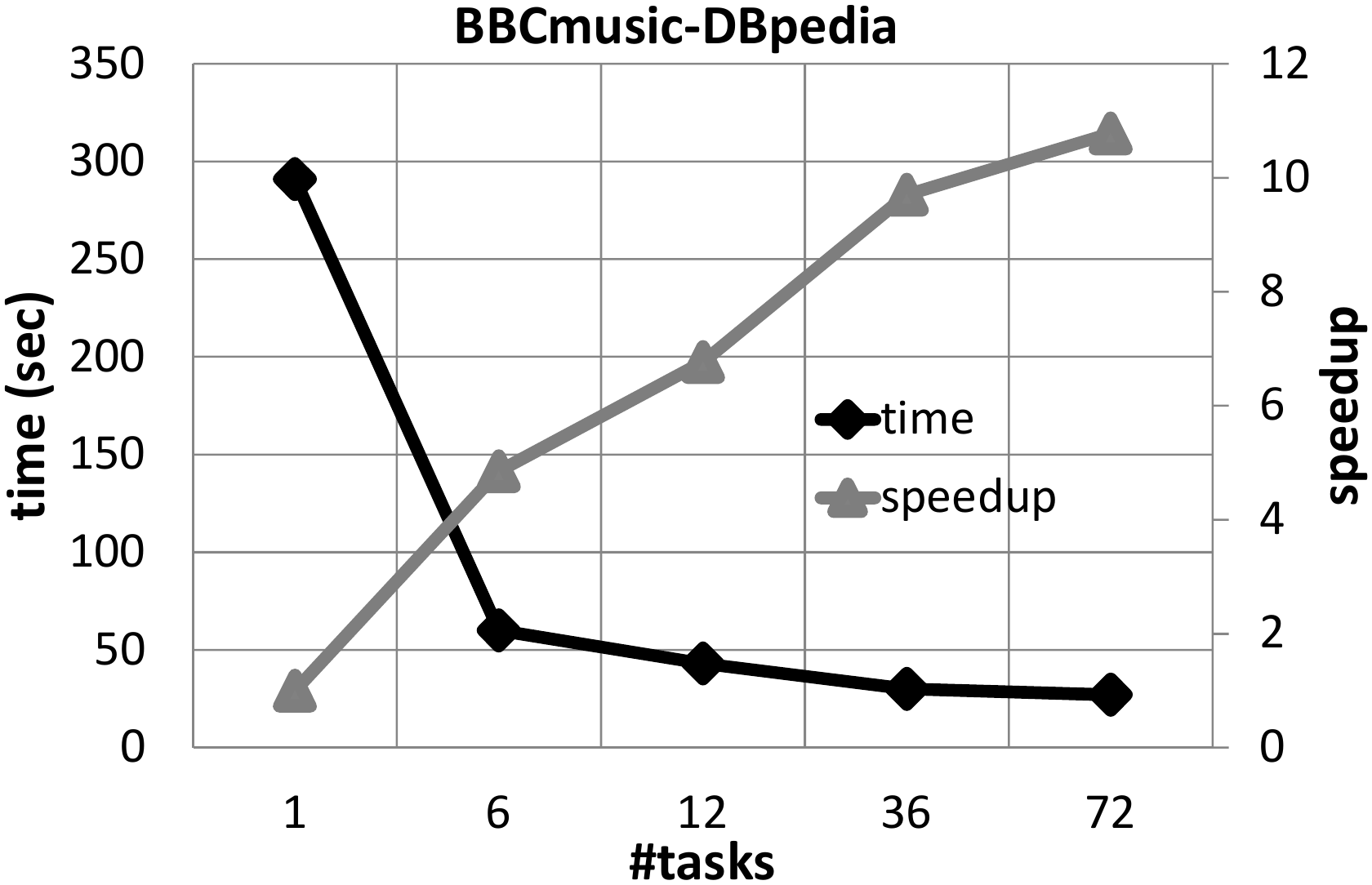}
	\includegraphics[width=0.245\linewidth]{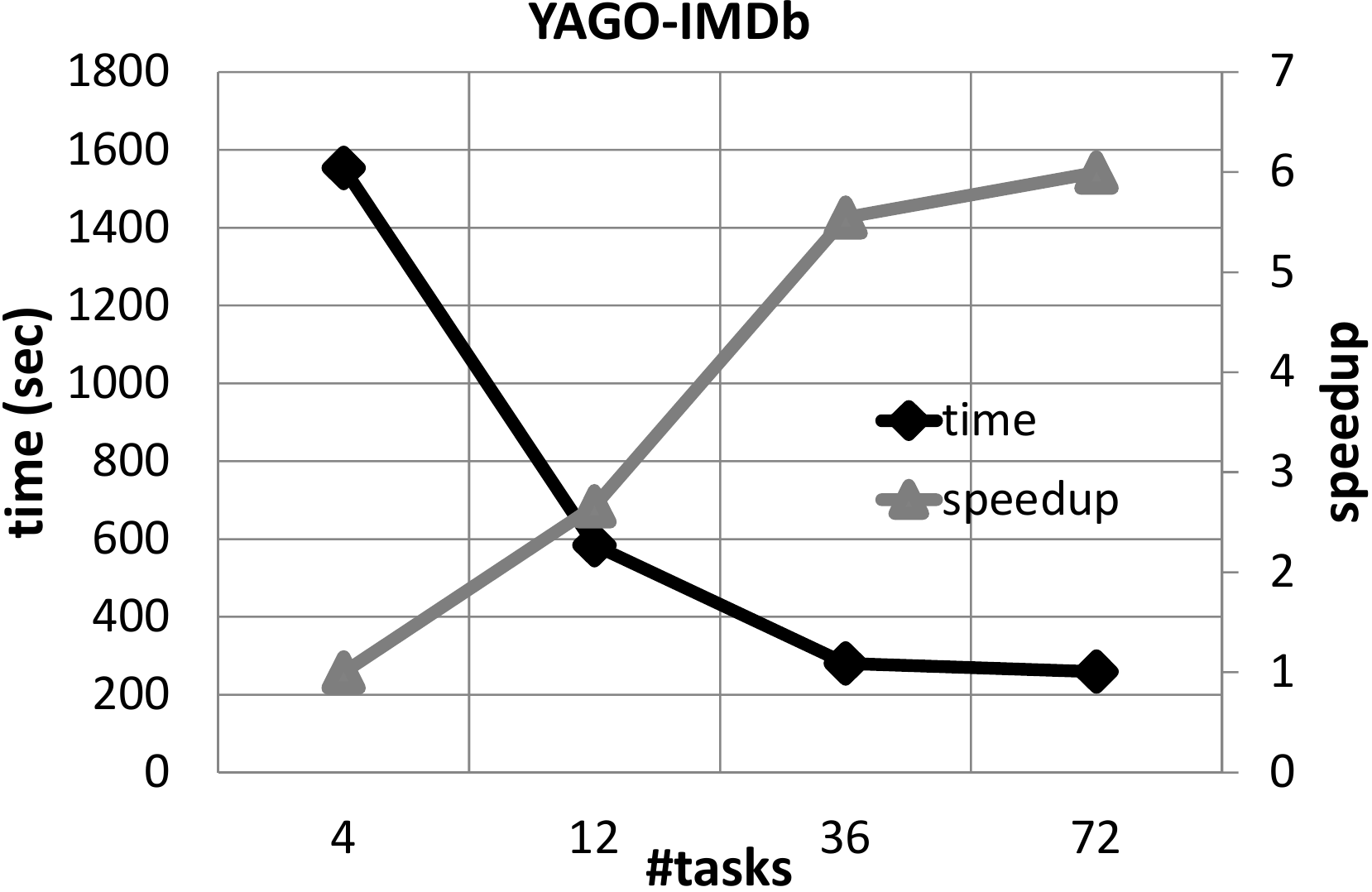}
	\caption{Scalability of matching in MinoanER w.r.t. running time (left vertical axis) and speedup (right vertical axis) as more cores are involved.}
	\vspace{-10pt}
	\label{fig:matching_scalability}
\end{figure*} 

$\bullet$ \textit{Name Matching Rule (\textsf{R1}).} This rule achieves both high precision ($>97\%$) and a decent recall ($>66\%$) in all cases. Hence, given no other matching evidence, \textsf{R1} alone yields good matching results, emphasizing precision, with only an insignificant number of its suggested matches being false positives. To illustrate the importance of this similarity evidence in real KBs, we have marked the matches with identical names in Figure~\ref{fig:similarityDistributions} as bordered points. Thus, we observe that matches may agree on their names, regardless of their value and neighbor similarity evidence. 

$\bullet$ \textit{Value Matching Rule (\textsf{R2}).} This rule is also very precise ($>90\%$ in all cases), but exhibits a lower recall ($>30\%$). Nevertheless, even this low recall is not negligible, especially when it complements the matches found from \textsf{R1}. In the case of strongly similar matches as in Restaurant, this rule alone can identify all the matches with perfect precision. 

$\bullet$ \textit{Rank Aggregation Matching Rule (\textsf{R3}).} The performance of this rule varies across the four datasets, as it relies on neighborhood evidence. For KBs with low value similarity (left part of Figure~\ref{fig:similarityDistributions}), this rule is the only solution for finding matches having no/different names. In BBCmusic-DBpedia and YAGO-IMDb, it has the highest contribution in recall and F1 of all matching rules, 
with the results for YAGO-IMDb being almost equivalent to those of Table~\ref{tab:matching_results} (YAGO-IMDb features the lowest value similarities in Figure~\ref{fig:similarityDistributions}). For KBs with medium value similarity (middle part of Figure~\ref{fig:similarityDistributions}), but not  enough to find matches with \textsf{R2}, aggregating neighbor with value similarity is very effective. In Rexa-DBLP, \textsf{R3} yields almost perfect results. Overall, \textsf{R3} is the matching rule with the greatest F1 in 3 out of 4 datasets.

$\bullet$ \textit{Reciprocity Matching Rule (\textsf{R4}).} Given that \textsf{R4} is a filtering rule, i.e., it does not add new results, we measure its contribution by running the full workflow without it. Its performance in Table~\ref{tab:heuristics} should be compared to the results in Table~\ref{tab:matching_results}. This comparison shows that this rule increases the precision of MinoanER, with a small, or no impact on recall. Specifically, it increases the precision of BBCmusic-DBpedia by $1.51\%$, while its recall is decreased by $1.38\%$, and in the case of YAGO-IMDb, it improves precision by $0.44\%$ with no impact on recall. This results in an increase of $0.04\%$ and $0.21\%$ in F1. 
Overall, \textsf{R4} is the weakest matching rule, yielding only a minor improvement in the results of MinoanER. 

\vspace{2pt}
\noindent
\textbf{Contribution of neighbors.} To evaluate the contribution of neighbor
evidence in the matching results, we have repeated
Algorithm~\ref{alg:matching_algorithm}, without the rule \textsf{R3}. 
Note that
this experiment is not the same as our baseline; here, we use all the other
rules, also operating on the pruned disjunctive blocking graph, while the
baseline does not use our rules and operates on the unpruned graph. The
results show that \textit{neighbor evidence plays a minor or even no role in KBs
with strongly similar entities}, i.e., Restaurant and Rexa-DBLP, \textit{while
having a bigger impact in KBs with nearly similar matches}, i.e.,
BBCmusic-DBpedia and YAGO-IMDb (see Figure~\ref{fig:similarityDistributions}). Specifically,
compared to the results of Table~\ref{tab:matching_results}, the use of neighbor
evidence improves precision by $2.22\%$ and recall by $3.19\%$ in
BBCmusic-DBpedia, while, in YAGO-IMDB, precision is improved by $2.97\%$ and
recall by $3.15\%$.

\subsection{Efficiency Evaluation}
\label{ssec:efficiency}

To evaluate the scalability of matching in MinoanER, we present
in Figure~\ref{fig:matching_scalability} the running times and speedup of
matching for each dataset, as we change the number of available CPU cores in our cluster, i.e., the number of tasks that can run at the same time. In each diagram, the left vertical axis shows the running time and the right vertical axis shows the speedup, as we increase the number of available cores (from 1 to 72) shown in the horizontal axis\footnote{We could not run MinoanER on the YAGO-IMDb dataset with only 1 core, due to limited space in a single machine, so we report its running time starting with a minimum of 4 cores. This means that the linear speedup for 72 tasks would not be 72, but 18 (72/4).}.  Across all experiments, we have kept the same total number of tasks, which was defined as the number of all cores in the cluster multiplied by a parallelism factor of 3, i.e., 3 tasks are assigned to each core, when all cores are available. This was to ensure that each task would require the same amount of resources (e.g., memory), regardless of the number of available cores. 

We observe that the running times decrease as more cores become available, and
this decrease is steeper when using a small number of cores. For example, resolving
Rexa-DBLP with 6 cores is 6 times faster than with 1 core, while it
is 10 times faster with 12 cores than with 1 core (top-right of
Figure~\ref{fig:matching_scalability}). We observe a sub-linear speedup
in all cases, which is expected when synchronization is required for different
steps (see Section~\ref{ssec:matching_implementation}). Though, the bigger datasets have a speedup closer to linear than
the smaller ones, because the Spark overhead is smaller with
respect to the overall running time in these cases. 
We have also measured the percentage of time spent for the matching phase (Algorithm \ref{alg:matching_algorithm}) compared to the total execution time of MinoanER. In Restaurant and Rexa-DBLP, matching takes 45\% of the total time, in BBCmusic-DBpedia 30\% and in YAGO-IMDb 20\%. Thus, in all cases, matching takes less than half the execution time, while it takes smaller percentage of time as the tasks get bigger.

It is not possible to directly compare the efficiency of MinoanER with the competitive tools of Table~\ref{tab:matching_results}; most of them are not publicly available, while the available ones do not support parallel execution using Spark. 
The running times reported in the original works are about sequential algorithms executed in machines with a different setting than ours. However, we can safely argue that our fixed-step process, as opposed to the data-iterative processes of existing works, boosts the efficiency of MinoanER at no cost in (or, in most cases, with even better) effectiveness. Indicatively, the running time of MinoanER for Rexa-DBLP was 3.5 minutes (it took PARIS 11 minutes on one of our cluster's servers - see Experimental Setup - for the same dataset), for BBCmusic-DBpedia it was 69 seconds (it took PARIS 3.5 minutes on one of our cluster's servers), while the running time for YAGO-IMDb was 28 minutes (SiGMa reports 70 minutes, and PARIS reports 51 hours). In small datasets like Restaurant, MinoanER can be slower than other tools, as Spark has a setup overhead, which is significant for such cases (it took MinoanER 27 seconds to run this dataset, while PARIS needed 6 seconds).

%% file: tables/datasets.tex
\begin{table}
\footnotesize
\centering
\caption{Dataset statistics.}
\vspace{-8pt}
\setlength{\tabcolsep}{4pt}
\begin{tabular}{| l | r | r | r | r |}
\cline{2-5}
\multicolumn{1}{c|}{} & \textbf{Restau-} & \textbf{Rexa-} & \textbf{BBCmusic-} & \textbf{YAGO-} \\
\multicolumn{1}{c|}{} & \textbf{rant} & \textbf{DBLP} & \textbf{DBpedia} & \textbf{IMDb} \\ \hline \hline
$\mathcal{E}_1$ entities & 339 & 18,492 & 58,793 & 5,208,100 \\ \hline
$\mathcal{E}_2$ entities & 2,256 & 2,650,832 & 256,602 & 5,328,774 \\ \hline
\hline 
$\mathcal{E}_1$ triples &1,130&87,519&456,304&27,547,595 \\ \hline
$\mathcal{E}_2$ triples &7,519&14,936,373&8,044,247&47,843,680 \\ \hline
\hline 
$\mathcal{E}_1$ av. tokens & 20.44 & 40.71 & 81.19 & 15.56 \\ \hline
$\mathcal{E}_2$ av. tokens & 20.61 & 59.24 & 324.75 & 12.49 \\ \hline
\hline 
$\mathcal{E}_1$/$\mathcal{E}_2$ attributes & 7 / 7 & 114 / 145 & 27 / 10,953 & 65 / 29\\ \hline
$\mathcal{E}_1$/$\mathcal{E}_2$ relations & 2 / 2 & 103 / 123 & 9 / 953 & 4 / 13\\ \hline
$\mathcal{E}_1$/$\mathcal{E}_2$ types & 3 / 3 & 4 / 11 & 4 / 59,801 & 11,767 / 15 \\ \hline
$\mathcal{E}_1$/$\mathcal{E}_2$ vocab. & 2 / 2 & 4 / 4  & 4 / 6 & 3 / 1 \\
\hline\hline
Matches & 89 & 1,309 & 22,770 & 56,683 \\ \hline 
\end{tabular}
\label{tab:matching_datasets}
\vspace{-5pt}
\end{table}

%% file: tables/blockStats.tex
\begin{table}
\footnotesize 
\centering
\caption{Block statistics. }
\vspace{-7pt}
\begin{tabular}{| l | r | r | r | r |}
\cline{2-5}
\multicolumn{1}{c|}{} & \textbf{Restaurant} & \textbf{Rexa-} & \textbf{BBCmusic-} & \textbf{YAGO-} \\
\multicolumn{1}{c|}{} & & \textbf{DBLP} & \textbf{DBpedia} & \textbf{IMDb} \\ 
\hline\hline
$|B_N|$ & 83 & 15,912 & 28,844 & 580,518\\
\hline
$|B_T|$ & 625 & 22,297 & 54,380 & 495,973\\
\hline\hline
$||B_N||$ & 83 & 6.71$\cdot$10$^7$ & 1.25$\cdot$10$^7$ & 6.59$\cdot$10$^6$\\
\hline
$||B_T||$ & 1.80$\cdot$10$^3$ & 6.54$\cdot$10$^8$ & 1.73$\cdot$10$^8$ & 2.28$\cdot$10$^{10}$\\
\hline
$|\mathcal{E}_1|\cdot|\mathcal{E}_2|$ & 7.65$\cdot$10$^5$ & 4.90$\cdot$10$^{10}$ & 1.51$\cdot$10$^{10}$ & 2.78$\cdot$10$^{13}$ \\
\hline\hline
Precision & 4.95 & 1.81$\cdot$10$^{-4}$ & 0.01 & 2.46$\cdot$10$^{-4}$ \\ \hline
Recall & 100.00 & 99.77 & 99.83 & 99.35 \\ \hline
F1 & 9.43 & 3.62$\cdot$10$^{-4}$ & 0.02 & 4.92$\cdot$10$^{-4}$ \\
\hline
\end{tabular}
\label{tab:blocks}
\vspace{-5pt}
\end{table}

%% file: tables/baselines.tex
\begin{table}[t]
\footnotesize  
\centering
\caption{Evaluation of MinoanER in comparison to the state-of-the-art methods and the heavily fine-tuned baseline, \textsf{BSL}.}
\vspace{-5pt}
\setlength{\tabcolsep}{4.2pt}
\begin{tabular}{| l | l | r | r | r | r |}
\cline{3-6}
\multicolumn{2}{c|}{} & \textbf{Restau-} & \textbf{Rexa-} & \textbf{BBCmusic-} & \textbf{YAGO-} \\
\multicolumn{2}{c|}{} & \textbf{rant} & \textbf{DBLP} & \textbf{DBpedia} & \textbf{IMDb} \\ \hline \hline
\multirow{3}{*}{SiGMa~\cite{DBLP:conf/kdd/Lacoste-JulienPDKGG13}} & Prec. & 99 & \textbf{97} & - & \textbf{98} \\ \cline{2-6}
 & Recall & 94 & 90 & - & 85 \\ \cline{2-6}
 & F1 & 97 & 94 & - & 91\\ \hline \hline

\multirow{3}{*}{LINDA~\cite{DBLP:conf/cikm/BohmMNW12}} & Prec. & \textbf{100} & - & - & - \\ \cline{2-6}
 & Recall & 63 & - & - & -\\ \cline{2-6}
 & F1 & 77 & - & - & -\\ \hline \hline

\multirow{3}{*}{RiMOM~\cite{DBLP:journals/tkde/LiTLL09}} & Prec. & 86 & 80 & - & - \\ \cline{2-6}
 & Recall & 77 & 72 & - & -\\ \cline{2-6}
 & F1 & 81 & 76 & - & -\\ \hline \hline

\multirow{3}{*}{PARIS~\cite{DBLP:journals/pvldb/SuchanekAS11}} & Prec. & 95 & 93.95 & 19.40 & 94 \\ \cline{2-6}
 & Recall & 88 & 89 & 0.29 & 90\\ \cline{2-6}
 & F1 & 91 & 91.41 & 0.51 & \textbf{92}\\ \hline \hline


\multirow{3}{*}{\textsf{BSL}} & Prec. & \textbf{100} & 96.57 & 85.20 &  11.68 \\ \cline{2-6}
 & Recall & \textbf{100} & 83.96 & 36.09 &  4.87\\ \cline{2-6}
 & F1 & \textbf{100} & 89.82 & 50.70 & 6.88\\ \hline \hline

\multirow{3}{*}{MinoanER} & Prec. & \textbf{100} & \textbf{96.74} & \textbf{91.44} & 91.02 \\ \cline{2-6}
 & Recall & \textbf{100} & \textbf{95.34} & \textbf{88.55} & \textbf{90.57} \\ \cline{2-6}
 & F1 & \textbf{100} & \textbf{96.04} & \textbf{89.97} & 90.79 \\ \hline
\end{tabular}
\label{tab:matching_results}
\vspace{-8pt}
\end{table}

%% file: tables/matchingRules.tex
\begin{table}
\footnotesize 
\centering
\caption{Evaluation of matching rules. }
\vspace{-5pt}
\begin{tabular}{| l | l | r | r | r | r |}
\cline{3-6}
\multicolumn{2}{c|}{} & \textbf{Restau-} & \textbf{Rexa-} & \textbf{BBCmusic-} & \textbf{YAGO-} \\
\multicolumn{2}{c|}{} & \textbf{rant} & \textbf{DBLP} & \textbf{DBpedia} & \textbf{IMDb} \\ \hline \hline
\multirow{3}{*}{R1} & Precision & \textbf{100} & \textbf{97.36} & \textbf{99.85} & 97.55 \\ \cline{2-6}
 & Recall & 68.54 & 87.47 & 66.11 & 66.53\\ \cline{2-6}
 & F1 & 81.33 & 92.15 & 79.55 & 79.11\\ \hline \hline

\multirow{3}{*}{R2} & Precision & \textbf{100} & 96.15 & 90.73 &  \textbf{98.02} \\ \cline{2-6}
 & Recall & \textbf{100} & 30.56 & 37.01 & 69.14 \\ \cline{2-6}
 & F1 & \textbf{100} & 46.38 & 52.66 & 81.08\\ \hline \hline

\multirow{3}{*}{R3} & Precision & 98.88 & 94.73 & 81.49 & 90.51 \\ \cline{2-6}
 & Recall & 98.88 & \textbf{94.73} & \textbf{81.49} & \textbf{90.50}\\ \cline{2-6}
 & F1 & 98.88 & \textbf{94.73} & \textbf{81.49} & \textbf{90.50}\\ \hline \hline

\multirow{3}{*}{$\neg$R4} & Precision & 100 & 96.03 & 89.93 & 90.58 \\ \cline{2-6}
 & Recall & 100 & 96.03 & 89.93 & 90.57 \\ \cline{2-6}
 & F1 & 100 & 96.03 & 89.93 & 90.58\\ \hline \hline
 
No & Precision & 100 & 96.59 & 89.22 & 88.05 \\ \cline{2-6}
Neigh- & Recall & 100 & 95.26 & 85.36 & 87.42 \\ \cline{2-6}
bors & F1 & 100 & 95.92 & 87.25 & 87.73 \\ \hline
\end{tabular}
\label{tab:heuristics}
\vspace{-8pt}
\end{table}

%% file: 7_conclusion.tex
\section{Conclusions}\label{sec:matching_conclusions}
In this paper, we have presented MinoanER, a fully automated, schema-agnostic and massively parallel framework for ER in the Web of Data. To resolve highly heterogeneous entities met in this context, MinoanER relies on schema-agnostic similarity metrics that consider both the content and the neighbors of entities. It exploits these metrics in a composite blocking scheme and conceptually builds a disjunctive blocking graph - a novel, comprehensive abstraction that captures all sources of similarity evidence. This graph of candidate matches is processed by a non-iterative algorithm that consists of four generic, schema-agnostic matching rules, with linear cost  to the number of entity descriptions and robust performance with respect to their internal configuration. 

The results show that neighbor evidence plays a minor role in KBs with strongly similar entities, such as Restaurant and Rexa-DBLP, while having a big impact in KBs with nearly similar entities, such as in BBCmusic-DBpedia and YAGO-IMDb. MinoanER achieves at least equivalent performance with state-of-the-art ER tools over KBs exhibiting low Variety, but outperforms them to a significant extent when matching KBs with high Variety. The employed matching rules (\textsf{R1}, \textsf{R2}, \textsf{R3}, \textsf{R4}) manage to cover a wide range of matches, as annotated in Figure~\ref{fig:similarityDistributions}, but there is still room for improvement, since the recall of blocking is better than that of matching.  As an improvement, we will investigate how to create an ensemble of matching rules and how to set the parameters of pruning candidate pairs dynamically, based on the local similarity distributions of each node's candidates.